\documentclass{article}

\usepackage{microtype}
\usepackage{graphicx}
\usepackage{subcaption}
\usepackage{booktabs} %

\usepackage{amsmath}
\usepackage{amsfonts}
\usepackage{enumerate}
\usepackage{grffile}  %

\usepackage{hyperref}

\usepackage[accepted]{icml2020}

\usepackage{amsthm}
\newtheorem{theorem}{Theorem}

\def\bx{{\mathbf x}}
\def\bz{{\mathbf z}}
\def\hatbx{{\mathbf{\hat x}}}
\def\hatbz{{\mathbf{\hat z}}}
\def\rate{{\mathcal R}}

\icmltitlerunning{Variational Bayesian Quantization}

\begin{document}

\twocolumn[
\icmltitle{Variational Bayesian Quantization}

\icmlsetsymbol{equal}{*}

\begin{icmlauthorlist}
\icmlauthor{Yibo Yang}{equal,to}
\icmlauthor{Robert Bamler}{equal,to}
\icmlauthor{Stephan Mandt}{to}
\end{icmlauthorlist}

\icmlaffiliation{to}{Department of Computer Science, University of California, Irvine}

\icmlcorrespondingauthor{Yibo Yang}{yibo.yang@uci.edu}
\icmlcorrespondingauthor{Robert Bamler}{rbamler@uci.edu}

\icmlkeywords{Machine Learning, ICML}

\vskip 0.3in
]

\printAffiliationsAndNotice{\icmlEqualContribution} %

\begin{abstract}
We propose a novel algorithm for quantizing continuous latent representations in trained models. Our approach applies to deep probabilistic models, such as variational autoencoders (VAEs), and enables both data and model compression. Unlike current end-to-end neural compression methods that cater the model to a fixed quantization scheme, 
our algorithm separates model design and training from quantization. 
Consequently, our algorithm enables ``plug-and-play'' compression with variable rate-distortion trade-off, using a \emph{single} trained model. 
Our algorithm can be seen as a novel extension of arithmetic coding to the continuous domain, and uses adaptive quantization accuracy based on estimates of posterior uncertainty.
Our experimental results demonstrate the importance of taking into account posterior uncertainties, and show that image compression with the proposed algorithm outperforms JPEG over a wide range of bit rates using only a single standard VAE.
Further experiments on Bayesian neural word embeddings demonstrate the versatility of the proposed method.
\end{abstract}

\section{Introduction}

Latent-variable models have become a mainstay of modern machine learning.
Scalable approximate Bayesian inference methods, in particular Black Box Variational Inference \citep{ranganath2014black, rezende2014stochastic}, have spurred the development of increasingly large and expressive probabilistic models, including deep generative probabilistic models such as variational autoencoders \citep{kingma2013auto} and Bayesian neural networks \citep{mackay1992practical, blundell2015weight}. 
~One natural application of deep latent variable modeling is data compression, and recent work has focused on end-to-end procedures that optimize a model for a particular compression objective.
Here, we study a related but different problem: given a \emph{trained} model, what is the best way to encode the information contained in its continuous latent variables?

As we show, our proposed solution provides a new ``plug-and-play'' approach to %
lossy compression of both \emph{data instances} (represented by local latent variables, e.g., in a VAE) as well as \emph{model parameters} (represented by global latent variables that serve as parameters of a Bayesian statistical model).
Our approach separates the compression task from model design and training, thus implementing variable-bitrate compression as an independent post-processing step in a wide class of existing latent variable models.

At the heart of our proposed method lies a novel quantization scheme that optimizes a rate-distortion trade-off by exploiting posterior uncertainty estimates.
Quantization is central to lossy compression, as continuous-valued data like natural images, videos, and distributed representations ultimately need to be discretized to a finite number of bits for digital storage or transmission. 
Lossy compression algorithms therefore typically find a discrete approximation of some semantic representation of the data, which is then encoded with a lossless compression method.

In classical lossy compression methods such as JPEG or MP3, the semantic representation is carefully designed to support compression at variable bitrates.
By contrast, state-of-the-art deep learning based approaches to lossy data compression \citep{balle2017end,balle2018variational,rippel2017realtime,mentzer2018conditional,lombardo2019deep} are trained to minimize a distortion metric at a fixed bitrate.
To support variable-bitrate compression in this approach, one has to train several models for different bitrates.
While training several models may be viable in many cases, a bigger issue is the increase in decoder size as the decoder has to store the parameters of not one but several deep neural networks for each bitrate setting.
In applications like video streaming under fluctuating connectivity, the decoder further has to load a new deep learning model into memory every time a change in bandwidth requires adjusting the bitrate.

By contrast, we propose a 
a quantization method for latent variable models
that decouples training from compression, and that enables variable-bitrate compression with a single model.
We generalize a classical entropy coding algorithm, Arithmetic Coding \citep{witten1987arithmetic, mackay2003information}, from the discrete to continuous domain.
Our proposed algorithm, Variational Bayesian Quantization, exploits posterior uncertainty estimates to automatically reduce the quantization accuracy of latent variables for which the model is uncertain anyway.
This strategy is analogous to the way humans communicate quantitative information.
For example, Wikipedia lists the population of Rome in 2017 with the specific number $2{,}879{,}728$.
By contrast, its population in the year $500$~AD is estimated by the round number $100{,}000$ because 
the high uncertainty would make a more precise number meaningless.
Our ablation studies show that this posterior-informed quantization scheme is crucial to obtaining competitive performance.

In detail, our contributions are as follows:
\begin{itemize}
    \item \textit{A new discretization scheme.}
    We present a novel approach to discretizing latent variables in a variational inference framework. Our approach generalizes arithmetic coding from discrete to continuous distributions and takes posterior uncertainty into account.
    \item \textit{Single-model compression at variable bitrates.}
    The decoupling of modeling and compression allows us to adjust the trade-off between bitrate and distortion in post-processing.
    This is in contrast to existing approaches to both data and model compression, which often require specially trained models for each bitrate.
    \item \textit{Automatic self-pruning.}
    Deep latent variable models often exhibit posterior collapse, i.e., the variational posterior collapses to the model prior. 
    In our approach, latent dimensions with collapsed posteriors require close to zero bits, thus don't require manual pruning.
    \item \textit{Competitive experimental performance.}
    We show that our method outperforms JPEG over a wide range of bitrates using only a single model. We also show that we can 
    successfully compress word embeddings with minimal loss, as evaluated on semantic reasoning task. 
\end{itemize}

The paper is structured as follows:
Section~\ref{sec:related} reviews related work in neural compression; Section~\ref{sec:method} proposes our Variational Bayesian Quantization algorithm. We give empirical results %
in Section~\ref{sec:experiments}, and conclude in 
Section~\ref{sec:conclusions}.
~ Section~\ref{sec:theory} provides additional theoretical insight about our method.

\section{Related Work}
\label{sec:related}

Compressing continuous-valued data is a classical problem in the signal processing community.
Typically, a distortion measure (often the squared error) and a source distribution are assumed, and the goal is to design a quantizer that optimizes the rate-distortion (R-D) performance \citep{lloyd1982least, Berger1972OptimumQA, Chou1989EntropyconstrainedVQ}.
Optimal vector quantization, although theoretically well-motivated \citep{gallager1968information}, is not tractable in high-dimensional spaces \citep{gersho2012vector} and not scalable in practice.
Therefore most classical lossy compression algorithms map data to a suitably designed semantic representation, in such a way that coordinate-wise scalar quantization can be fruitfully applied.

Recent machine-learning-based data compression methods learn such hand-designed representation from data, but similar to classical methods, most such ML methods directly take quantization into account in the generative model design or training.
Various approaches have been proposed to approximate the non-differentiable quantization operation during training, such as stochastic binarization \citep{toderici2015variable,toderici2017full}, additive uniform noise \citep{balle2017end,balle2018variational,habibian2019video}, or other differentiable approximation \citep{agustsson2017soft, theis2017lossy, mentzer2018conditional, rippel2017realtime}; many such schemes result in quantization with a uniformly-spaced grid, with the exception of \citep{agustsson2017soft}, which optimizes for quantization grid points. \citet{yang2020improving} considers optimal quantization at compression time, but assumes a fixed quantization scheme of \citep{balle2017end} during training.

We depart from such approaches by treating quantization as a post-processing step decoupled from model design and training. 
Crucial to our approach is
a new quantization scheme that automatically adapts to different length scales in the representation space based on posterior uncertainty estimates.
To our best knowledge, 
the only prior work that uses posterior uncertainty for compression is in the context of bits-back coding \citep{honkela2004variational, townsend2019practical}, but these works focus on lossless compression, with the recent exception of \citep{yang2020improving}. %

Most existing neural image compression methods require training a separate machine learning model for each desired bitrate setting \citep{balle2017end, balle2018variational, mentzer2018conditional,theis2017lossy,lombardo2019deep}.
In fact, \citet{alemi2018fixing} showed that any particular fitted VAE model only targets one specific point on the rate-distortion curve.
Our approach has the same benefit of variable-bitrate single-model compression as methods based on recurrent VAEs \citep{gregor2016conceptual, toderici2015variable, toderici2017full, johnston2018improved}; but unlike these methods, which use dedicated model architecture for progressive image reconstruction, we instead focus more broadly on quantizing latent representations in a given generative model, designed and trained for specific application purposes (possibly other than compression, e.g., modeling complex scientific observations).

\section{Posterior-Informed Variable-Bitrate Compression}
\label{sec:method}

We now propose an algorithm for quantizing latent variables in trained models. 
After describing the problem setup and assumptions (Subsection~\ref{sec:models}), we briefly review 
Arithmetic Coding (Subection~\ref{sec:arithmetic-coding}).
Subsection~\ref{sec:bayesian-ac} describes our proposed lossy compression algorithm, which generalizes Arithmetic Coding to the continuous domain.

\subsection{Problem Setup}
\label{sec:models}

\paragraph{Generative Model and Variational Inference.}
We consider a wide class of generative probabilistic models with data~$\bx$ and unknown (or ``latent'') variables $\bz\in\mathbb R^K$ from some continuous latent space with dimension~$K$.
The generative model is defined by a joint probability distribution,
\begin{align}\label{eq:joint-distribution}
    p(\bx,\bz) = p(\bz) \, p(\bx|\bz)
\end{align}
with a prior $p(\bz)$ and a likelihood $p(\bx|\bz)$. Although our presentation focuses on unsupervised representation learning, our framework also captures the supervised setup.%
\footnote{For supervised learning with labels~$y$, we would consider a conditional generative model $p(y,\bz|\bx) =  p(y|\bz,\bx)\, p(\bz)$ with conditional likelihood $p(y|\bz,\bx)$, where $\bz$ are the model parameters, treated as a Bayesian latent variable with associated prior $p(\bz)$.}

Our proposed compression method uses~$\bz$ as a proxy to describe the data~$\bx$.
This requires ``solving'' Eq.~\ref{eq:joint-distribution} for~$\bz$ given~$\bx$, i.e., inferring the posterior
$p(\bz|\bx) = p(\bx,\bz) / \int\! p(\bx,\bz)\, d\bz$.
Since exact Bayesian inference is often intractable,  %
we resort to Variational Inference (VI)~\citep{jordan1999introduction,blei2017variational, zhang2019advances}, which approximates the posterior by a so-called variational distribution $q_\phi(\bz|\bx)$ by minimizing the Kullback-Leibler divergence $D_\text{KL}(q_\phi(\bz|\bx) \,||\, p(\bz|\bx))$ over a set of variational parameters~$\phi$.

\paragraph{Factorization Assumptions.}\label{par:factorization-assumptions}
We assume that both the prior $p(\bz)$ and the variational distribution $q_\phi(\bz|\bx)$ are fully factorized (mean-field assumption).
For concreteness, our examples use a Gaussian variational distribution.
Thus,
\begin{align}%
    p(\bz) &= \textstyle{\prod}_{i=1}^K p(z_i); \label{eq:factorized-prior}
    \qquad\text{and} \\
    q_\phi(\bz|\bx) &= \textstyle{\prod}_{i=1}^K \mathcal N(z_i; \mu_i(\bx), \sigma^2_i(\bx)), \label{eq:factorized-q}
\end{align}
where $p(z_i)$ is a prior for the $i$\textsuperscript{th} component of~$\bz$, and the means~$\mu_i$ and standard deviations~$\sigma_i$ together comprise the variational parameters~$\phi$ over which VI optimizes.%
\footnote{These parameters are often amortized by a neural network (in which case $\mu_i$ and $\sigma_i$ depend on~$\bx$), but don't have to (in which case $\mu_i$ and $\sigma_i$ do not depend on $\bx$ and are directly optimized).}

Prominently, the model class defined by Eqs.~\ref{eq:joint-distribution}-\ref{eq:factorized-q} includes variational autoencoders (VAEs)~\citep{KW2014} for data compression, but we stress that the class is much wider, capturing also Bayesian neural nets \citep{mackay2003information}, 
probabilistic word embeddings \citep{barkan2017bayesian,bamler2017dynamic}, matrix factorization \citep{mnih2008probabilistic}, and topic models \citep{blei2003latent}.

\paragraph{Protocol Overview.}
We consider two parties in communication, a sender and a receiver.
In the case of \emph{data} compression, both parties have access to the model, but only the sender has access to the data point~$\bx$, which it uses to fit a variational distribution $q_\phi(\bz|\bx)$. %
It then uses the algorithm proposed below to select a latent variable vector~$\hatbz$ that has high probability under $q_\phi$, and that can be encoded into a compressed bitstring, which gets transmitted to the receiver. 
The receiver  losslessly decodes the compressed bitstring back into~$\hatbz$ and uses the likelihood $p(\bx|\hatbz)$ to generate a reconstructed data point $\hatbx$, typically setting $\hatbx = \arg \max_\bx p(\bx|\hatbz)$.
 In the case of \emph{model} compression, the sender infers a distribution $q_\phi(\bz|\bx)$ over model parameters $\bz$ given training data $\bx$, and uses our algorithm to select a suitable vector $\hatbz$ of quantized model parameters. The receiver receives~$\hatbz$ and uses it to reconstruct the model.
\label{compression-protocal}

The rest of this section describes how the proposed algorithm selects~$\hatbz$ and encodes it into a compressed bitstring.

\subsection{Background: Arithmetic Coding}
\label{sec:arithmetic-coding}

Our quantization algorithm, introduced in Section~\ref{sec:bayesian-ac} below, is inspired by a lossless compression algorithm, \emph{arithmetic coding} (AC) \citep{witten1987arithmetic, mackay2003information}, which we generalize from discrete data to the continuous space of latent variables $\bz\in\mathbb R^K$.
To get there, we first review the main idea of AC that our proposed algorithm borrows.

AC is an instance of so-called entropy coding.
It uniquely maps messages~$m\in \mathcal M$ from a discrete set~$\mathcal M$ to a compressed bitstring of some length~$\rate_m$ (the ``bitrate'').
Entropy coding exploits prior knowledge of the distribution $p(m)$ of messages to map probable messages to short bitstrings while spending more bits on improbable messages.
This way, entropy coding algorithms aim to minimize the expected rate $\mathbb E_{p(m)}[\mathcal R_m]$.
For lossless compression, the expected rate has a fundamental lower bound, the entroy $H=\mathbb E_{p(m)}[h(m)]$, where $h(m)=-\log_2 p(m)$ is the Shannon information content of~$m$.
AC provides near optimal lossless compression as it maps each message~$m\in\mathcal M$ to a bitstring of length $\rate_m=\lceil h(m)\rceil$, where $\lceil\cdot\rceil$ denotes the ceiling function.

\begin{figure}[t]
\begin{center}
\includegraphics[width=\columnwidth]{figs/method/ac-encoding.pdf}
\end{center}
\caption{Comparison of Arithmetic Coding (AC, left) and VBQ (right, proposed).
Both methods use a prior CDF (orange) to map nonuniformly distributed data to a number $\xi\sim \mathcal U(0,1)$, and both require an uncertainty region for truncation.}
\label{fig:ac}
\end{figure}

AC is usually discussed in the context of streaming compression where $m$ is a sequence of symbols from a finite alphabet, as AC improves on this task over the more widely known Huffman coding~\citep{huffman1952method}.
In our work, we focus on a different aspect of AC: its use of a cumulative probability distribution function to map a nonuniformly distributed random variable $m \sim p(m)$ to a number~$\xi$ that is nearly uniformly distributed over the interval $[0,1)$.

Figure~\ref{fig:ac} (left) illustrates AC for a binomial-distributed message $m\in\{0,\ldots,10\}$ (the number of `heads' in a sequence of ten coin flips).
The solid and dashed orange lines show the left and right sided cumulative distribution function,%
\footnote{If $m$ is a sequence of symbols, $F_<$ and $F_\leq$ are defined by lexicographical order and can be constructed in a streaming manner.}
$F_<(m):=\sum_{m'<m} p(m')$ and $F_\leq(m):=\sum_{m'\leq m} p(m')$,
respectively.
They define a partitioning of the interval $[0,1)$ (vertical axis in Figure~\ref{fig:ac} (left)) into pairwise disjoint subintervals $\mathcal I_m := \big[F_<(m), F_\leq(m)\big) $ (orange squares).
Since the intervals $\mathcal I_m$ are disjoint for all $m\in\mathcal M$, any number $\xi\in\mathcal I_m$ uniquely identifies a given message~$m$.
AC picks such a number~$\hat\xi \in \mathcal I_m$ and encodes it into a string of bits~$b_\kappa$, $\kappa \in \{0, \ldots, \rate_m\}$ by writing it in binary representation,
\begin{align}\label{eq:xi-m}
    \hat\xi = {\left(0.b_1b_2\ldots b_{\rate_{m}}\right)}_2
    \quad\text{with bits $b_\kappa\in\{0,1\}\; \forall \kappa$.}
\end{align}

Since any $\xi\in\mathcal I_m$ may be used to identify the message~$m$, we can interpret the interval~$\mathcal I_m$ as an \emph{uncertainty region} in $\xi$-space.
AC picks the number $\hat\xi\in\mathcal I_m$ with the shortest binary representation.
This requires at most $\lceil h(m)\rceil$ bits because the numbers~$\xi$ that can be represented by Eq.~\ref{eq:xi-m} with $\rate_m=\lceil h(m)\rceil$ form a uniform grid with spacing $2^{-\rate_{m}}=2^{-\lceil h(m)\rceil}$, which is at most as wide as the size of the interval, $|\mathcal I_{m}|=p(m)=2^{-h(m)}$.
The red arrows in Figure~\ref{fig:ac} (left) illustrate how AC would encode the message $m=7$ in the toy example into the bitstring ``$111$''.
Decoding works in the opposite direction and maps $\hat\xi$ back to~$m$.

In the next section, we generalize AC to the continuous domain.
As we will show, the concept of an ``uncertainty region'' in $\xi$-space becomes again crucial.

\subsection{Variational Bayesian Quantization}
\label{sec:bayesian-ac}

We now present our proposed algorithm, Variational Bayesian Quantization (VBQ), a novel quantization method for lossy compression that is inspired by AC but that operates on \emph{continuous} latent variables $\bz\in\mathbb R^K$.
Similar to AC, VBQ exploits knowledge of a prior probability distribution $p(\bz)$ in combination with a (soft) uncertainty region to encode probable values of~$\bz$ into short bitstrings.

\paragraph{From Intervals to Distributions.}
The main ideas that VBQ borrows from AC are as follows: (1)~the use of a cumulative distribution function to map a non-uniformly distributed random variable to a uniformly distributed random variable $\xi$ over the interval $(0,1)$, and (2)~the use of an ``uncertainty region" to select a number from this interval to encode the message with as few bits as possible. While AC uses an interval~$\mathcal I_m$ with hard boundaries, VBQ softens this uncertainty region 
by considering posterior uncertainty.

We consider a single continuous latent variable $z_i\in\mathbb R$ with arbitrary prior $p(z_i)$.
The cumulative (CDF) of the prior, %
\begin{align} %
    F(z_i) := \int_{-\infty}^{z_i} p(z_i')\, d z_i'\, ,
\end{align}
is shown in orange in Figure~\ref{fig:ac} (right).
It maps $z_i \sim p(z_i)$ to $\xi_i \sim \mathcal U(0,1)$. 
In contrast to the discrete case discussed in Section~\ref{sec:arithmetic-coding}, where the prior CDF maps each message~$m$ to an entire interval $\mathcal I_m$, note that the CDF of a continuous random variable maps real numbers to real numbers.

Since $\xi_i\sim \mathcal U(0,1)$ is almost surely an irrational number, its binary representation is infinitely long, and thus has to be truncated.
We find an optimal truncation by generalizing the idea of the uncertainty region~$\mathcal I_m$ to the continuous space:
we consider the posterior uncertainty in $z_i$-space and map it to $\xi_i$-space.
Approximating the posterior $p(z_i|\bx)$ by the variational distribution $q(z_i|\bx) := \mathcal N(z_i;\mu_i(\bx),\sigma_i^2(\bx))$, see Eq.~\ref{eq:factorized-q}, we thus consider the function
\begin{align}\label{eq:def-g}
    g(\xi_i) := q(F^{-1}(\xi_i) \,|\, \bx).
\end{align}
Here, $F^{-1}$ is the inverse CDF (the quantile function), which maps $\xi_i$ back to~$z_i$.
Note that $g$ is not a normalized probability distribution, as Eq.~\ref{eq:def-g} deliberately does not include the Jacobian $\nabla_{\!\xi_i}F^{-1}(\xi_i)$ because the final objective will be to maximize $q_\phi(\bz|\bx)$ at a single point (see Eq.~\ref{eq:log-q-log-g} below).

\paragraph{Intuition.}
The solid and dashed purple curves in figure Figure~$\ref{fig:ac}$ (right) plot $q(z_i|\bx)$ and $g(\xi_i)$ on the horizontal and vertical axis, respectively.
The red arrows illustrate how a finite uncertainty region $\mu_i(\bx) \pm \sigma_i(\bx)$ in $z_i$-space is mapped to a finite width of~$g$ in $\xi_i$-space.
VBQ finds a quantile~$\hat\xi_i$ that has high value under $g$ while at the same time having a short binary representation.
The two purple arrowheads on the vertical axis point to two viable candidates, $\hat \xi_i=\frac78$ and $\hat \xi_i=\frac34$, that both lie within the uncertainty region.
The choice between these two points poses a rate-distortion trade-off:
while $\frac78 \equiv {(0.111)}_2$ has higher value under~$g$ (i.e., it identifies a point $\hat z_i= F^{-1}(\frac78)$ with higher approximate posterior probability $q(\hat z_i|\bx)$), the alternative $\hat\xi_i=\frac34 \equiv {(0.11)}_2$ can be encoded in fewer bits.

\paragraph{Optimizing the Rate-Distortion Trade-Off.}
Rather than considering a hard uncertainty region, VBQ simply tries to find a point $\boldsymbol\xi\equiv(\xi_i)_{i=1}^K$ that identifies latent variables $\bz\equiv (z_i)_{i=1}^K$ with high probability under the variational distribution $q_\phi(\bz|\bx)$ while being expressible in few bits.
We thus express $\log q_\phi(\bz|\bx)$ in terms of the coordinates $\xi_i = F(z_i)$ using Eq.~\ref{eq:factorized-q},
\begin{align}\label{eq:log-q-log-g}
    \log q_\phi(\bz|\bx)
    &=-\sum_{i=1}^K \frac{\big( F^{-1}( \xi_i) - \mu_i(\bx)\big)^2}{2\,\sigma_i^2(\bx)} + \text{cnst.}
\end{align}

For each dimension~$i$, we restrict the quantile~$\xi_i \in (0, 1)$ to the set of code points $\hat\xi_i$ that can be represented in binary via Eq.~\ref{eq:xi-m} with a finite but arbitrary bitlength $\rate(\hat\xi_i)$.
We define the total bitlength $\rate(\boldsymbol{\hat\xi}):=\sum_{i=1}^K\rate(\hat\xi_i)$, i.e., the length of the concatenation of all codes $\hat\xi_i$, $i\in\{1,\ldots,K\}$ neglecting, for now, an overhead for delimiters (see below). 
Using a rate penalty parameter $\lambda>0$ that is shared across all dimensions~$i$, we minimize the rate-distortion objective
\begin{align}\label{eq:rd-objective}
      \mathcal L_\lambda(\boldsymbol{\hat\xi}|\bx)
      &= -\log q_\phi(\hatbz|\bx) + \lambda \rate(\boldsymbol{\hat\xi}) \\
      &= \sum_{i=1}^K \left[ \frac{\big(F^{-1}(\hat \xi_i) - \mu_i(\bx)\big)^2}{2\,\sigma_i^2(\bx)} + \lambda \mathcal R(\hat \xi_i) \right] + \text{cnst}. \nonumber
\end{align}
The optimization thus decouples across all latent dimensions~$i$, and can be solved efficiently and in parallel by minimizing the $K$ independent objective functions
\begin{align}\label{eq:objective-single-dim}
    \ell_\lambda(\hat\xi_i|\bx)=  \big(F^{-1}(\hat \xi_i) - \mu_i(\bx)\big)^2 + 2\lambda\,\sigma_i^2(\bx) \, \mathcal R(\hat \xi_i).
\end{align}

\begin{algorithm}[tb]
   \caption{Rate-Distortion Optimization for Dimension $i$}
   \label{alg:scalar-rd-optimization}
\begin{algorithmic}
    \STATE \begin{tabular}{@{}l@{~}l@{}}
    {\bfseries Input:} & Prior CDF $F(z_i)$,
        rate penalty $\lambda>0$, \\
     & variational mode  $\mu_i(\bx)$ and variance $\sigma^2_i(\bx)$. \\
    {\bfseries Output:} & Optimal code point $\hat\xi_i^*\equiv (0.b_1b_2\ldots b_{\rate(\hat\xi_i^*)})_2$.
    \end{tabular}
    \STATE Evaluate $\xi_i^\dagger \gets F(\mu_i(\bx))$.
    \STATE Initialize $r\gets0,\quad \hat\xi_i^* \gets \text{null},\quad
    \ell^* \leftarrow \infty$. \vspace{1pt}
    \REPEAT
        \STATE Update $r\gets r+1$
        \STATE Set $\hat\xi_i^{r,\text{left}} \leftarrow 2^{-r} \lfloor 2^{r} \xi_i^\dagger\rfloor,\quad
        \hat\xi_i^{r,\text{right}} \gets 2^{-r} \lceil 2^{r} \xi_i^\dagger\rceil$.
        \IF{$\hat\xi_i^{r,\text{left}}\neq0$ \textbf{and} $\ell_{\lambda}(\hat\xi_i^{r,\text{left}} \,|\,\bx) < \ell^*$  }
          \STATE Update $\hat\xi_i^* \gets \hat\xi_i^{r,\text{left}},\quad
          \ell^* \leftarrow \ell_{\lambda}(\hat\xi_i^{r,\text{left}} \,|\,\bx)$.
        \ENDIF
        \IF{$\hat\xi_i^{r,\text{right}}\neq1$ \textbf{and} $\ell_{\lambda}(\hat\xi_i^{r,\text{right}} \,|\,\bx) < \ell^*$  }
          \STATE Update $
          \hat\xi_i^* \gets \hat\xi_i^{r,\text{right}},\quad
          \ell^* \leftarrow \ell_{\lambda}(\hat\xi_i^{r,\text{right}} \,|\,\bx)$.
        \ENDIF
        
    \UNTIL{$\log g(\xi_i^\dagger) -\log g(\hat\xi_i^*) < \lambda (r+1-\rate(\hat\xi_i^*))$.}
\end{algorithmic}
\end{algorithm}

Although the bitlength $R(\hat \xi_i)$ is discontinuous (it counts the number of binary digits, see Eq.~\ref{eq:xi-m}), $\ell_\lambda(\hat\xi_i|\bx)$ can be efficiently minimized over~$\hat\xi_i$ using Algorithm \ref{alg:scalar-rd-optimization}.
The algorithm iterates over all rates $r\in\{1,2,\ldots\}$ and searches for the code point~$\hat\xi_i^*$ that minimizes $\ell(\hat\xi_i|\bx)$.
For each~$r$, the algorithm only needs to consider the two code points $\hat\xi_i^{r,\text{left}} \leq \xi_i^\dagger$ and $\hat\xi_i^{r,\text{right}} \geq \xi_i^\dagger$ with rate at most $r$ that enclose the optimum $\xi_i^\dagger:= F(\mu_i(\bx))$ and are closest to it; these two code points can be easily computed in constant time.
The iteration terminates as soon as the maximally possible remaining increase in $\log q(z_i|\bx)=\log g(\xi_i)$ is smaller than the minimum penalty for an increasing bitlength (in practice, the iteration rarely exceeds $r\approx 8$).

\paragraph{Encoding.}
After finding the optimal code points $(\hat\xi_i^*)_{i=1}^K$, they have to be encoded into a single bitstring.
Simply concatenating the binary representations (Eq.~\ref{eq:xi-m}) of all $\hat\xi_i^*$ would be ambiguous due to their variable lengths $\rate(\hat\xi_i^*)$
(see detailed discussion in the Supplementary Material).
Instead, we treat the code points as symbols from a discrete vocabulary and encode them via lossless entropy coding, e.g., Arithmetic Coding.
The entropy coder requires a probabilistic model over all code points; here we simply use their empirical distribution.
When using our method for model compression, this empirical distribution has to be transmitted to the receiver as additional header information that counts towards the total bitrate.
For data compression, by contrast, we obtain the empirical distribution of code points on training data %
and include it in the decoder.

\begin{figure}[t]
\begin{center}
\includegraphics[width=\columnwidth]{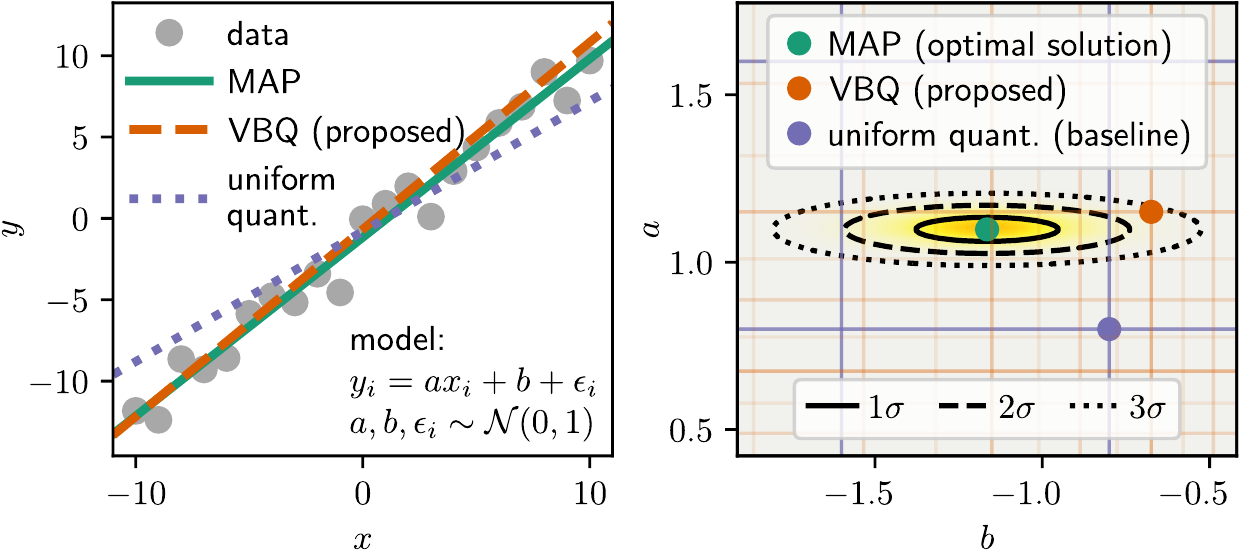}
\end{center}
\caption{Effect of an anisotropic posterior distribution on quantization.
Left: linear regression model with optimal fit (green) and fits of models with quantized parameters (orange, purple).
Right: posterior distribution and quantized model parameters following two different quantization schemes.
Although both quantized models are equally far away from the optimal solution (green dot), VBQ (orange) fits the data better because it takes the anisotropy of the posterior into account.
}
\label{fig:anisotropy}
\end{figure}

\paragraph{Discussion.}\label{par:method-discussion}
The proposed %
algorithm adjusts the accuracy for each latent variable~$z_i$ based on two factors:
(i)~a \textit{global} rate setting $\lambda$ that is shared across all dimensions~$i$; and
(ii)~a per-dimension posterior uncertainty estimate $\sigma_i(\bx)$.
Point~(i) allows tuning the rate-distortion trade-off whereas (ii)~takes the anisotropy of the latent space into account.

Figure~\ref{fig:anisotropy} illustrates the effect of anisotropy in latent space.
The right panel plots the posterior of a toy Bayesian linear regression model $y=ax+b$ (see left panel) with only two latent variables $\bz\equiv(a,b)$.
Due to the elongated shape of the posterior, VBQ uses a higher accuracy for~$a$ than for~$b$. 
As a result, the algorithm finds a quantization~$\hatbz$ (orange dot in right panel) that is closer to the optimal (MAP) solution (green dot) along the $a$-axis than along the $b$-axis.

The purple dot in Figure~\ref{fig:anisotropy} (right) compares to a more common quantization method, which simply
rounds the MAP solution to the nearest point (which is then entropy coded) from a fixed grid with spacing $\delta>0$. 
We tuned $\delta$ so that the resulting quantized model parameters (purple dot) have the same distance to the optimum as our proposed solution (orange dot).
Despite the equal distance to the optimum, VBQ finds model parameters with higher posterior probability. The resulting model fits the data better (left panel).

This concludes the description of the proposed Variational Bayesian Quantization algorithm.
In the next section, we analyze the algorithm's behaviour experimentally and demonstrate its performance for variable-bitrate compression on both word embeddings and images.

\section{Experiments}
\label{sec:experiments}

We tested our approach on two very different domains: word embeddings and images.
For word embeddings, we measured the performance drop on a semantic reasoning task due to lossy compression.
Our proposed VBQ method significantly improves model performance over uniform discretization and compression with either Arithmetic Coding (AC), gzip, bzip2, or lzma at equal bitrate.
For image compression, we show that a single standard VAE, compressed with VBQ,  outperforms JPEG and other baselines at a wide range of bitrates, both quantitatively and visually.

\subsection{Compressing Word Embeddings}
\label{sec:word-embeddings}

\begin{figure}[t]
\centering
 \includegraphics[width=\columnwidth]{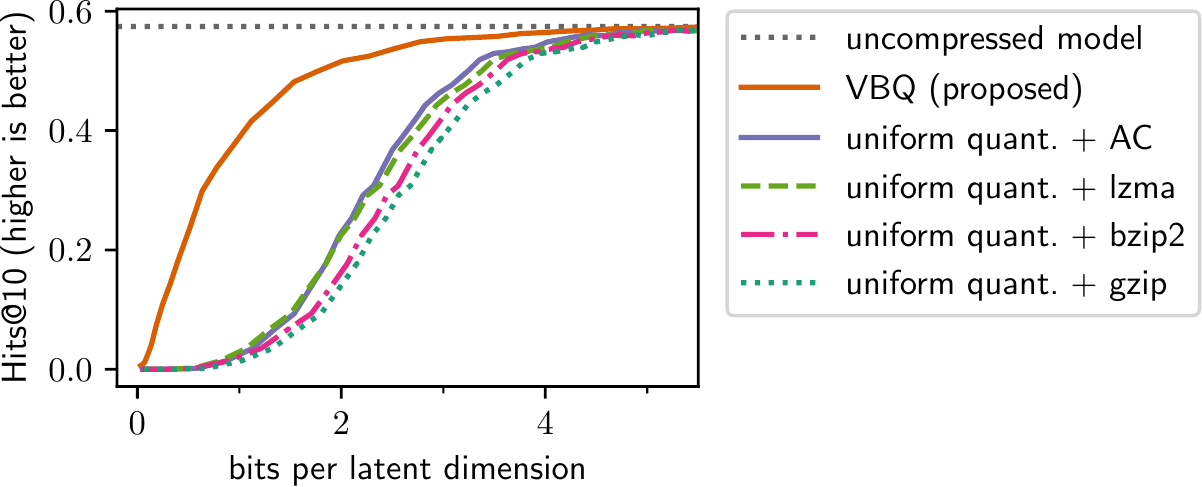}
\caption{Performance of compressed word embeddings on a standard semantic and syntactic reasoning task~\cite{mikolov2013efficient}.
VBQ (orange, proposed) leads to much smaller file sizes at equal model performance over a wide range of performances.}
\label{fig:word-embeddings}
\end{figure}

We consider the Bayesian Skip-gram model for neural word embeddings~\citep{barkan2017bayesian}, a probabilistic generative formulation of word2vec~\cite{mikolov2013distributed} which interprets word and context embedding vectors as latent variables and associates them with Gaussian approximate posterior distributions. Point estimating the latent variables would result in classical word2vec. 
Even though the model was not specifically designed or trained with model compression taken into consideration, the proposed algorithm can successfully compress it in post-processing.

\paragraph{Experiment Setup.}
We implemented the Black Box VI version of the Bayesian Skip-gram model proposed in \citep{bamler2017dynamic},%
\footnote{See Supplementary Material for hyperparameters. Our code is available at \url{https://github.com/mandt-lab/vbq}.}
and trained the model on books published between $1980$ and $2008$ from the Google Books corpus~\citep{michel2011quantitative}, following the preprocessing described in \citep{bamler2017dynamic} with a vocabulary of $V=100{,}000$ words and embedding dimension $d=100$.

In the trained model, we observed that the distribution of posterior modes $\mu_{w,j}$ across all words~$w$ and all dimensions~$j$ of the embedding space was quite different from the prior.
To improve the bitrate of our method, we used an ``empirical prior'' for encoding that is shared across all~$w$ and~$j$; we chose a Gaussian $\mathcal N(0,\sigma_0^2)$ where $\sigma_0^2$ is the empirical variance of all variational means $(\mu_{w,j})_{w=1,\ldots, V;\, j=1,\ldots, d}$. 

We compare our method's performance to a baseline that quantizes to a uniform grid and then uses the empirical distribution of quantized coordinates for lossless entropy coding.
We also compare to uniform quantization baselines that replace the entropy coding step with the standard compression libraries gzip, bzip2, and lzma.
These methods are not restricted by a factorized distribution of code points and could therefore detect and exploit correlations between quantized code points across words or dimensions.

\begin{figure}[t!]
\begin{center}
    \includegraphics[width=\columnwidth]{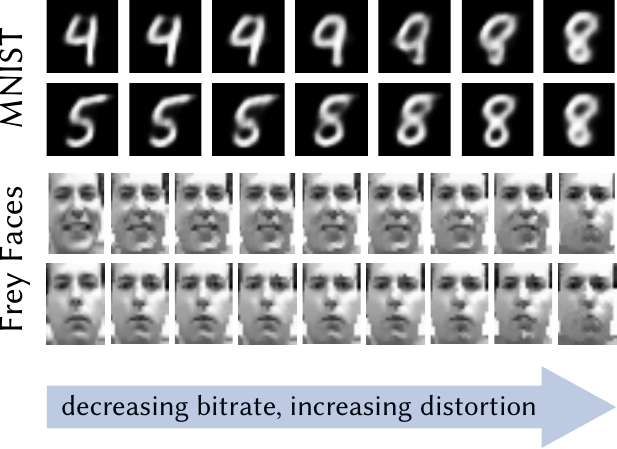}
\end{center}
\caption{Qualitative behavior of our proposed VBQ algorithm on two data sets of small-scale images (MNIST and Frey Faces).
With decreasing bitrate, the method starts to confuse the encoded object with a generic one (encoded by the median of the prior $p(\bz)$).
}
\label{fig:mnist-series}
\end{figure}

We evaluate performance on the semantic and syntactic reasoning task proposed in \citep{mikolov2013efficient}, a popular dataset of semantic relations like ``Japan~:~yen = Russia~:~ruble'' and syntactic relations like ``amazing~:~amazingly = lucky~:~luckily'', where the goal is to predict the last word given the first three words.
We report Hits@10, i.e., the fraction of challenges for which the compressed model ranks the correct prediction among the top ten.

\paragraph{Results.}
Figure~\ref{fig:word-embeddings} shows the model performance on the semantic and syntactic reasoning tasks as a function of compression rate.
Our proposed VBQ significantly outperforms all baselines and reaches the same Hits@10 at less than half the bitrate over a wide range.%
\footnote{
The uncompressed model performance (dotted gray line in Figure~\ref{fig:word-embeddings}) is not state of the art.
This is not a shortcoming of the compression method but merely of the model, and can be attributed to the smaller vocabulary and training set used compared to \citep{mikolov2013distributed} due to hardware constraints.}

\subsection{Image Compression}
\label{sec:image-compression}

\begin{figure*}[t!]
\centering
    \begin{subfigure}[h]{0.19\textwidth}
    {\includegraphics[width=\textwidth]{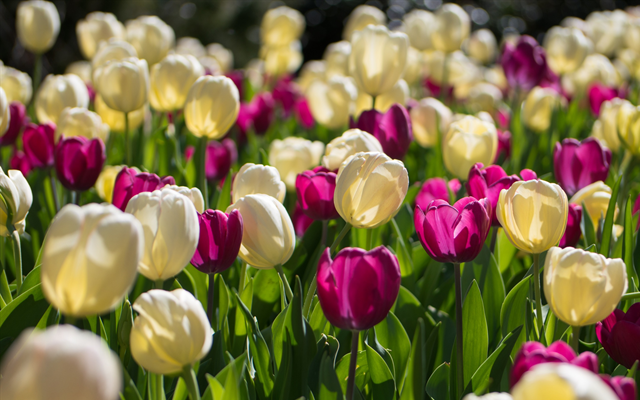}}\caption{Original \newline}
    \end{subfigure}
    \begin{subfigure}[h]{0.19\textwidth}
    {\includegraphics[width=\textwidth]{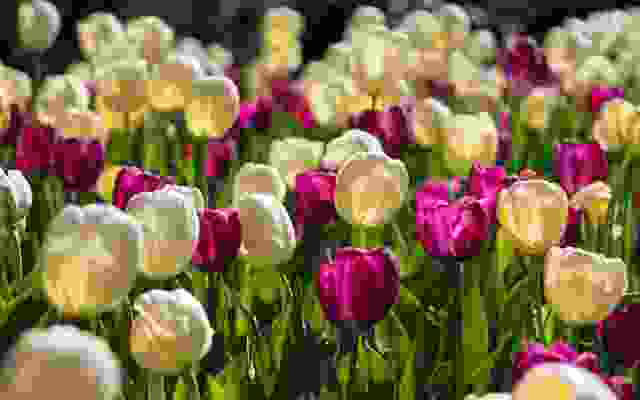}}\caption{JPEG \newline \hspace*{1em} MS-SSIM=$0.813$}
    \end{subfigure}
    \begin{subfigure}[h]{0.19\textwidth}
    {\includegraphics[width=\textwidth]{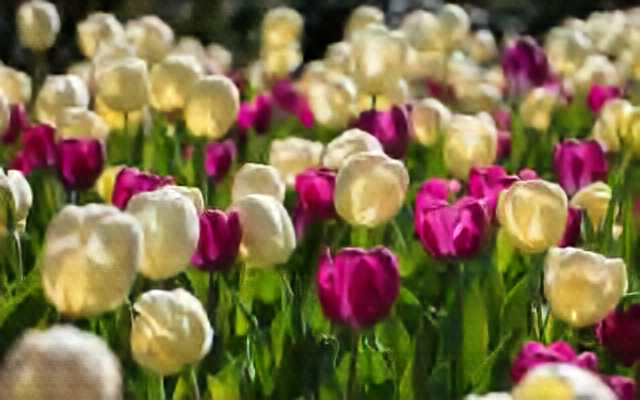}}\caption{VBQ \newline \hspace*{1em} MS-SSIM=$0.933$}
    \end{subfigure}
    \begin{subfigure}[h]{0.19\textwidth}
    {\includegraphics[width=\textwidth]{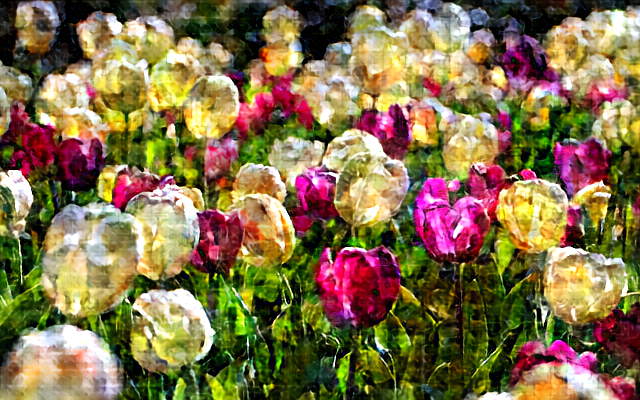}}\caption{Uniform grid \newline \hspace*{1em} MS-SSIM=$0.723$}
    \end{subfigure}
    \begin{subfigure}[h]{0.19\textwidth}
        {\includegraphics[width=\textwidth]{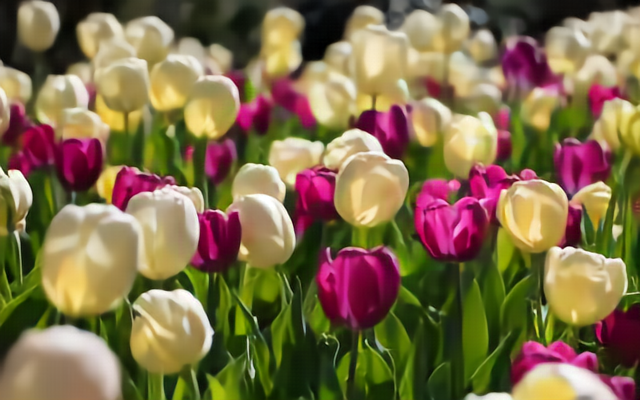}}\caption{Ball{\'e} et al. \newline \hspace*{1em} MS-SSIM=$0.958$}
    \end{subfigure}

\caption{Image reconstructions at matching bitrate (0.24 bits per pixel).
VBQ~(c; proposed) outperforms AC with uniform quantization~(d) and JPEG~(b) and is comparable to the approach by \citep{balle2017end}~(e) despite using a model that is not optimized for this specific bitrate. Uniform quantization here used a modified version of the VAE in Figure~\ref{fig:kodak-rd}, 
using an additional conv layer with smaller dimensions to reduce the bitrate down to 0.24
(this was not possible in the original model even with the largest possible grid spacing).
}
\label{fig:example-reconstructions}
\end{figure*}

While Section~\ref{sec:word-embeddings} demonstrated the proposed VBQ method for model compression, we now apply the same method to data compression using a variational autoencoder (VAE).
We first provide 
qualitative insight on small-scale images, and then
quantitative results on full resolution color images.

\paragraph{Model.}
For simplicity, we consider regular VAEs with a standard normal prior
and Gaussian variational posterior. 
The generative network parameterizes a factorized categorical or Gaussian likelihood model in experiments in Sec. \ref{sec:toy} or \ref{sec:kodak},
respectively. Network architectures are described below and in more detail in Supplementary Material.

\paragraph{Baselines.}

We consider the following baselines:
\begin{itemize}
\setlength\itemsep{-.1em}
    \item \emph{Uniform quantization:} for a given image~$\bx$, we quantize each dimension of the posterior mean vector $\boldsymbol{\mu}(\bx)$ to a uniform grid.
    We report the bitrate for encoding the resulting quantized latent representation via standard entropy coding (e.g., arithmetic coding).
    Entropy coding requires prior knowledge of the probabilities of each grid point.
    Here, we use the empirical frequencies of grid points over a subset of the training set;
    \item \emph{$k$-means quantization}: similar to ``uniform quantization'', but with the placement of grid points optimized via $k$-means clustering on a subset of the training data;
    \item Quantization with \emph{generalized Lloyd algorithm}: similar to above, but the grid points are optimized using generalized Lloyd algorithm \citep{Chou1989EntropyconstrainedVQ}, a widely-used state-of-the-art classical quantization method;
    \item \emph{JPEG:} we used the libjpeg implementation packaged with the Python Pillow library, using default configurations (e.g., 4:2:0 subsampling), and we adjust the quality parameter to vary the rate-distortion trade-off;
    \item \emph{Deep learning baseline:} we compare to \citet{balle2017end}, who directly optimized for the rate and distortion, training a separate model for each point on the R-D curve. 
    In our large-scale experiment, we adopte their 
    model architecture,
    so their performance essentially represents the end-to-end optimized performance upper bound for our method (which uses a single model). 
\end{itemize}

\subsubsection{Qualitative Analysis on Toy Datasets}
\label{sec:toy}

We trained a VAE on the MNIST dataset and the Frey Faces dataset, using 5 and 4-dimensional latent spaces, respectively. See Supplemental Material for experimental details.

Figure \ref{fig:mnist-series} shows example image reconstructions from our VBQ algorithm with increasing $\lambda$, and thus decreasing bitrate. The right-most column is the extreme case $\lambda \to \infty$, resulting in the shortest possible bistring encoding $\hat \xi_i = (0.1)_2 = \frac12$ (i.e., $\hat z_i$ being the median of the prior $p(z_i)$) for every dimension $i$.
As the bitrate decreases (as $\rate(\boldsymbol{\hat\xi}) \to 0$), our method gradually ``confuses" the original image with a generic image (roughly in the center of the embedding space), while preserving approximately the same level of sharpness.
This is in contrast to JPEG which typically introduces blocky and/or pixel-level artifacts at lower bitrates.

\subsubsection{Full-Resolution Color Images} \label{sec:kodak}
We apply our VBQ method to a VAE trained on color images, and 
obtain practical image compression performance rivaling JPEG, while outperforming baselines that ignore posterior uncertainty and directly quantize latent variables.

\paragraph{Model and Dataset.}
The inference and generative networks of our VAE are identical to the analysis and synthesis networks of \citet{balle2017end}, using 3 layers of 256 filters each in a convolutional architecture.  
We used a diagonal Gaussian likelihood model, whose mean is computed by the generative net and the variance $\sigma^2$ is fixed as a hyper-parameter, similar to a $\beta$-VAE \citep{higgins2016beta} approach ($\sigma^2$ was tuned to 0.001 to ensure the VAE achieved overall good R-D trade-off; see \citep{alemi2018fixing}).
We trained the model on the same subset of the ImageNet dataset as used in \citep{balle2017end}. 
We evaluated performance on the standard Kodak \citep{kodak-cd} dataset,
a separate set of 24 uncompressed color images.
As in the word embedding experiment, we also observed that using an empirical prior for our method improved the bitrate; for this,
we used the flexible density model of \citet{balle2018variational}, 
fitting a different distribution 
per latent channel, on samples of posterior means $\boldsymbol\mu$ (treating spatial dimensions as i.i.d.).

\begin{figure}[t]
\centering
 \centering
 \includegraphics[width=0.5\textwidth]{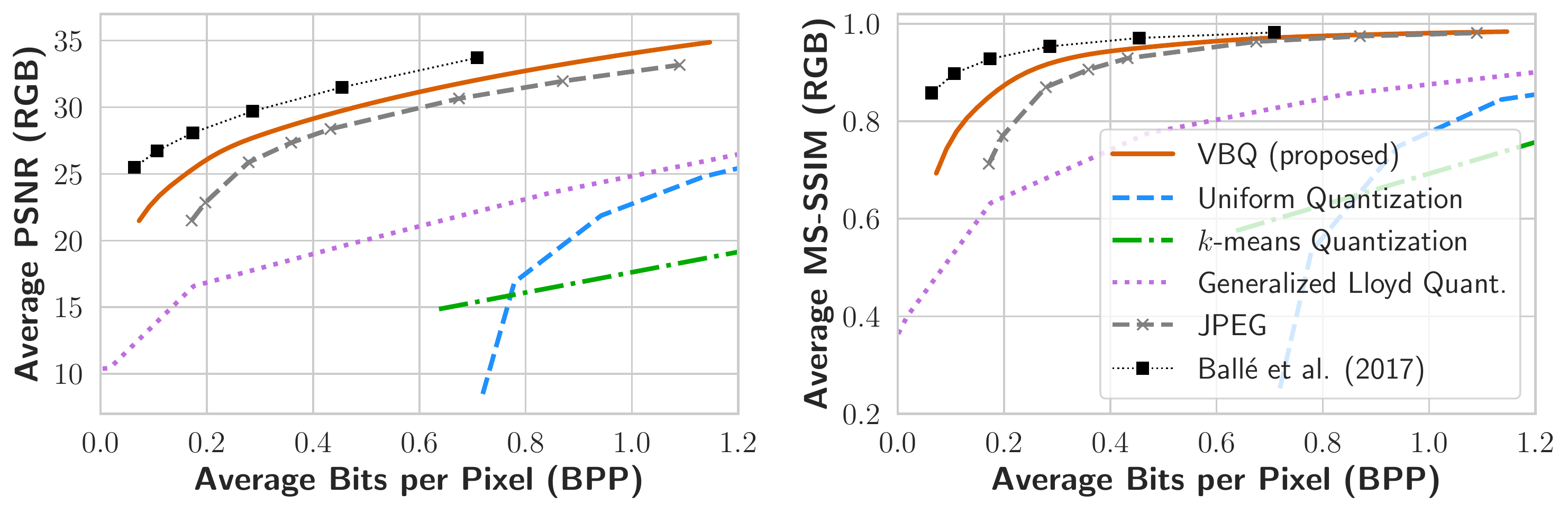}
\caption{Aggregate rate-distortion performance on the Kodak dataset (higher is better).
VBQ (blue, proposed) outperforms JPEG for all tested bitrates with a single model.
By contrast, \citep{balle2017end} (black squares) relies on individually optimized models for each bitrate that all have to be included in the decoder.
}\label{fig:kodak-rd}
\end{figure}

\paragraph{Results.}
As common in image compression work, we measure distortion between original and compressed images under two quality metrics (the higher the better): Peak Signal-to-Noise ratio (PSNR) and MS-SSIM \citep{wang2003multiscale}, over all RGB channels.
Figure~\ref{fig:kodak-rd} shows rate-distortion performance.
We averaged both the bits per pixel (BPP) and quality measure over all images in the Kodak dataset for each~$\lambda$ 
(we got similar results by averaging only over the quality metrics for fixed bitrates via interpolation).

We found that our method generally produced images with higher quality, both in terms of PSNR and perceptual quality, compared to JPEG and uniform quantization. Similar to  \citep{balle2017end}, our method avoids unpleasant artifacts, and introduces blurriness at low bitrate. See Figure \ref{fig:example-reconstructions} for example image reconstructions. For more examples and R-D curves on individual images, see Supplementary Material. 

Although our results fall short of the end-to-end optimized rate-distortion performance of \citet{balle2017end}, it is worth emphasizing that our method operates anywhere on the R-D curve with a \emph{single} trained VAE model, unlike \citet{balle2017end}
, which
requires costly optimization and storage of individual models for each point on the R-D curve. 
On the other hand, as with any quantization method, the reconstruction quality of VBQ is upper-bounded by that of the full-precision latents; e.g., evaluating on uncompressed latents gives a PSNR upper-bound of about 38.9 in Figure~\ref{fig:kodak-rd}.

\begin{figure}[t]
\begin{center}
\includegraphics[width=\columnwidth]{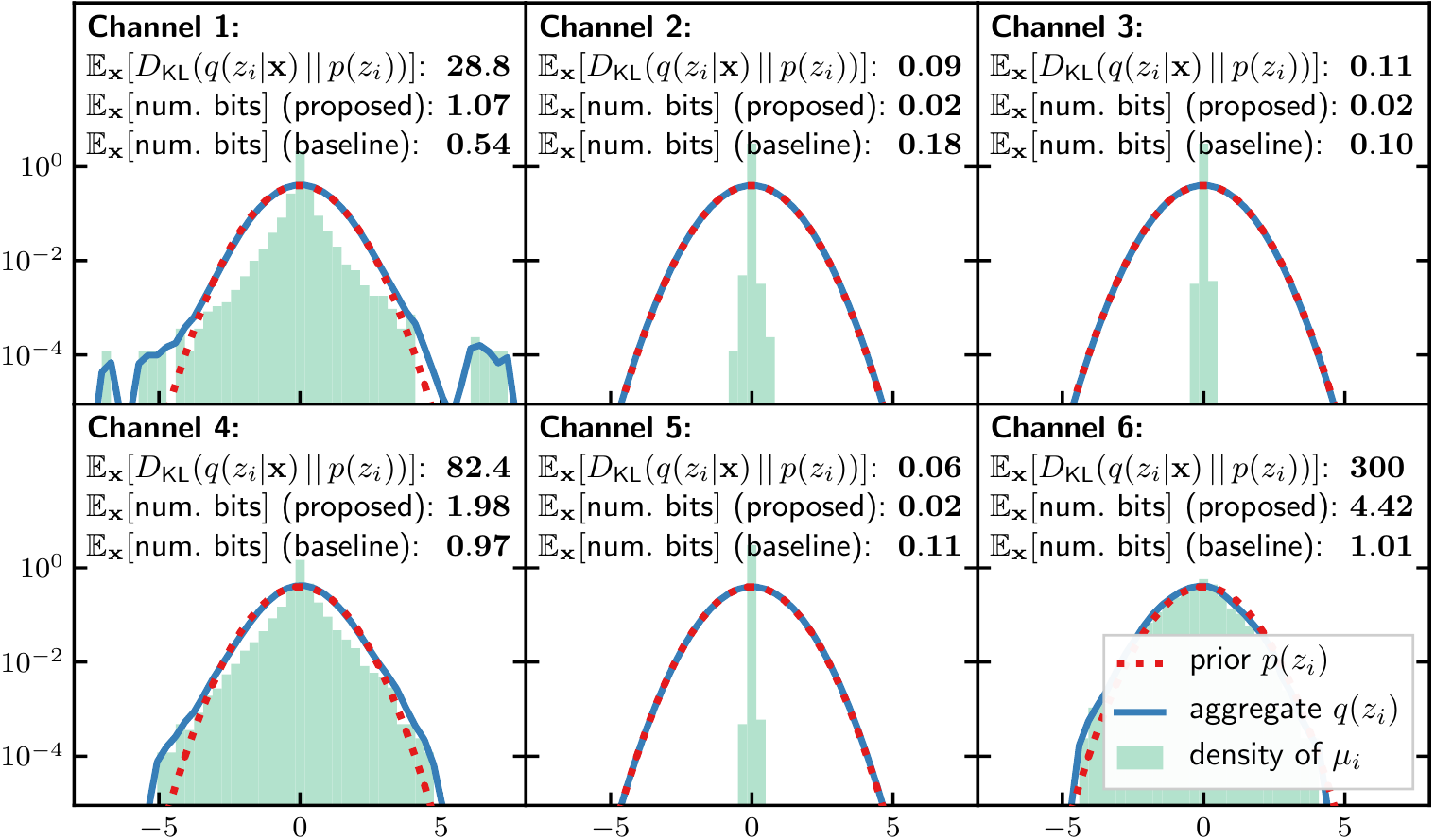}
\end{center}
\caption{Variational posteriors and the encoding cost for the first~6 latent channels of an image-compression VAE trained with high~$\beta$ setting.
``Baseline'' refers to uniform quantization.
The proposed VBQ method wastes fewer bits on channels that exhibit posterior collapse (channels 2, 3, and 5) than the baseline method.
It instead spends more bits on channels without posterior collapse.
}
\label{fig:posterior_collapse}
\end{figure}

\paragraph{Indifference to Posterior Collapse.}
A known issue in deep generative models such as VAEs is
posterior collapse, where the model 
ignores some subset of latent variables, 
whose
variational posterior distributions collapse to closely match the prior. 
Such collapsed dimensions constitute an overhead in conventional neural compression approaches, 
which is often dealt with by a pruning step.
One curious consequence of our approach is that it automatically spends close to zero bits encoding the collapsed latent dimensions.

As an illustration, we trained a VAE as used in the color image compression experiment with a high $\beta$ setting to purposefully induce posterior collapse, and examine the average number of bits spent on various latent channels. %
Figure \ref{fig:posterior_collapse} shows the prior $p(z_i)$, aggregated (approximate) posterior $q(z_i) :=\mathbb{E}_\bx[q(z_i|\bx)]$, and histograms of posterior means $\mu_i(\bx)$
for the first six channels of the VAE; all the quantities were averaged over an image batch and across latent spatial dimensions.
We observe that channels 2, 3, and 5 appear to exhibit posterior collapse, as the aggregated posteriors closely match the prior while the 
posterior means tightly cluster at zero;
this is also reflected by low average KL-divergence between the variational posterior $q(z_i|\bx)$ and prior $p(z_i)$, see text inside each panel.
We observe that, for these collapsed channels, our method spends fewer bits on average than uniform quantization (baseline) at the same total bitrate, and more bits 
instead on channels 1, 4, and 6, which do not exhibit posterior collapse.
The explanation is that a collapsed posterior has unusually high variance $\sigma_i^2(\bx)$, causing our model to refrain from long code words due to the high penalty $\propto \sigma_i^2(\bx)$ per bitrate $\rate(\hat\xi_i)$ in Eq.~\ref{eq:objective-single-dim}.

\section{Conclusions}
\label{sec:conclusions}

We proposed a novel algorithm for discretizing latent representations in trained latent variables, with applications to both data and model compression. 
Our proposed ``Variational Bayesian Quantization'' (VBQ) algorithm automatically adapts encoding accuracy to posterior uncertainty estimates, and is applicable to arbitrary probabilistic generative models with factorized prior and mean-field variational posterior distributions.
As our approach separates quantization from model design and training, it enables ``plug-\&-play'' compression at variable bitrate with pretrained models. 

We showed the empirical effectiveness of our proposed VBQ method for both model and data compression.
For model compression, VBQ retained significantly higher task performance of a trained word embedding model than other  methods that compress in post-processing.
For data compression, we showed that VBQ can outperform JPEG over a wide range of bitrates with a single trained standard VAE.
Given its versatility, we believe that VBQ holds promise for compressing Bayesian neural networks, especially in applications that demand rate-distortion trade-offs.
 Lastly, as VBQ relies on 
 accurate posterior approximation
 , its rate-distortion performance provides a new metric for quantitative evaluation of approximate Bayesian inference methods.

\section{Theoretical Considerations}
\label{sec:theory}
Here, we provide additional theoretical insights into the proposed VBQ method based on reviewer feedback.

\paragraph{Dense Quantization Grid.}
Section~\ref{sec:bayesian-ac} describes the proposed VBQ algorithm in terms of quantiles~$\xi_i$.
While quantiles simplify the discussion, it is also instructive to consider what effectively happens directly in representation space.

\begin{figure}[t]
    \vskip 0.06in
    \centering
    \includegraphics[width=0.49\textwidth]{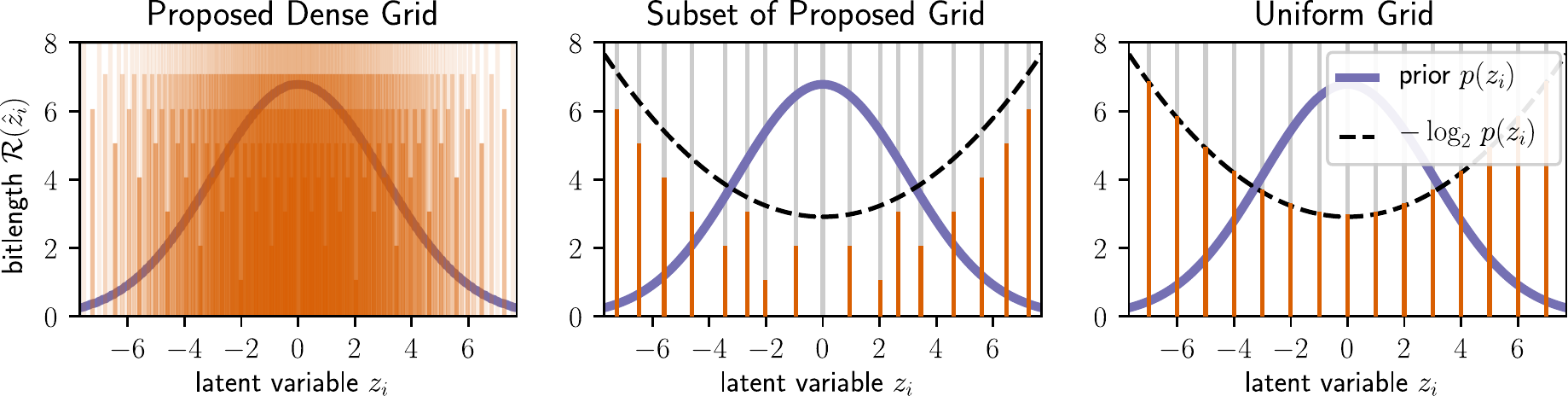}
    \vspace{-0.2in}
    \caption{Dense quantization grid of VBQ (left) and a subset of it (center) that resembles the more common uniform grid (right).}
    \label{fig:grids}
\end{figure}

In the space of representations~$\bz$, VBQ optimizes over a dense quantization grid, shown in Figure~\ref{fig:grids} (left) for a single dimension~$z_i$.
Each code point $\hat{\xi_i} \in (0, 1)$, i.e., each quantile with a finite-length binary representation, defines a grid point $\hat{z_i} = F^{-1}(\hat \xi_i)$.
The height of each orange bar in the figure shows the bitlength $\mathcal R(\hat z_i)$.
Interestingly, the grid places many points with small $\mathcal R(\hat z_i)$ in regions of high prior probability density (purple curve), while regions of low prior probability contain only grid points with large $\mathcal R(\hat z_i)$.
This observation can be formalized as follows.

\begin{theorem} \label{thm:almost-equidistant}
    For a latent dimension~$z_i$ with prior $p(z_i)$ and for any interval $\mathcal I\subset \mathbb R$, there exists a grid point $\hat z_i \in \mathcal I$ whose bitlength is bounded by the information content of~$\mathcal I$, i.e.,
    $\mathcal R(\hat z_i) \leq - \log_2 P(\mathcal I)$ with $P(\mathcal I) = \int_{\mathcal I} p(z_i)\, dz_i$.
\end{theorem}
\begin{proof}
    $\mathcal R(\hat z_i)$ is the number of nontrivial bits in the quantile $\hat\xi_i := F(\hat z_i)$.
    For example, the quantiles $\frac14={(0.01)}_2$, $\frac12={(0.1)}_2$, and $\frac34={(0.11)}_2$ each have at most one nontrivial bit (since the initial~``$0.$'' and terminal~``$1$'' are considered trivial), and they are equally spaced with spacing $\frac14 = 2^{-2}$.
    Incrementing the number of bits by one divides the spacing in half.
    Thus, for any $r\in\mathbb N$, the quantiles with at most~$r$ nontrivial bits are equally spaced with spacing $2^{-r-1}$.
    The prior CDF~$F$ maps any interval $\mathcal I\subset \mathbb R$ to an interval $F(\mathcal I)\subset(0,1)$ of size $|F(\mathcal I)| = P(\mathcal I)$.
    Since $P(\mathcal I) > 2^{-r-1}$ for $r=\lfloor -\log_2 P(\mathcal I)\rfloor$, the interval~$\mathcal I$ contains at least one grid point $\hat z_i$ with $\mathcal R(\hat z_i) \leq r$.
\end{proof}

\paragraph{Delimination Overhead.}
The above theorem provides an upper bound on the bitrate for a single coordinate~$z_i$.
Encoding a high dimensional latent representation~$\bz$ 
requires some overhead for delimiting the individual coordinates.

We provide a theoretical upper bound on this overhead for a severely restricted variant of VBQ that does not have access to posterior uncertainty estimates and that operates only on a subset of the proposed grid.
Specifically, we pick only one grid point for each interval $\mathcal I_n = [n-\frac12, n+\frac12)$ with $n\in\mathbb Z$.
According to Theorem~\ref{thm:almost-equidistant}, there always exists a grid point $\hat z_i \in \mathcal I_n$ with $\mathcal R(\hat z_i) \leq -\log_2 P(\mathcal I_n)$.
Thus, the resulting subset of the grid (Figure~\ref{fig:grids}, center) resembles the more commonly used uniform grid (Figure~\ref{fig:grids} right), whose bitrate under standard entropy coding is the information content.

If we restrict VBQ to this sparse subset of the dense grid, then the algorithm collapses to a method known as Shannon-Fano-Elias coding \citep{cover2012elements}, whose overhead over the information theoretically minimal bitrate is known to be at most one bit per dimension.
The full VBQ algorithm has more freedom: it exploits posterior uncertainty estimates to reduce accuracy where it is not needed, thus saving bits and improving compression performance.

\section*{Acknowledgements}
We thank Johannes Ball{\'e} for helping us reproduce the results in prior work \citep{balle2017end}. Yibo Yang acknowledges funding from the Hasso Plattner Foundation. This material is based upon work supported by the Defense Advanced Research Projects Agency (DARPA) under Contract No. HR001120C0021.  Any opinions, findings and conclusions or recommendations expressed in this material are those of the author(s) and do not necessarily reflect the views of the Defense Advanced Research Projects Agency (DARPA). Furthermore, this work was supported by the National Science Foundation under Grants NSF-1928718, NSF-2003237, and by Qualcomm.

\bibliography{references}

\begin{thebibliography}{44}
\providecommand{\natexlab}[1]{#1}
\providecommand{\url}[1]{\texttt{#1}}
\expandafter\ifx\csname urlstyle\endcsname\relax
  \providecommand{\doi}[1]{doi: #1}\else
  \providecommand{\doi}{doi: \begingroup \urlstyle{rm}\Url}\fi

\bibitem[Agustsson et~al.(2017)Agustsson, Mentzer, Tschannen, Cavigelli,
  Timofte, Benini, and Gool]{agustsson2017soft}
Agustsson, E., Mentzer, F., Tschannen, M., Cavigelli, L., Timofte, R., Benini,
  L., and Gool, L.~V.
\newblock Soft-to-hard vector quantization for end-to-end learning compressible
  representations.
\newblock In \emph{Advances in Neural Information Processing Systems}, 2017.

\bibitem[Alemi et~al.(2018)Alemi, Poole, Fischer, Dillon, Saurous, and
  Murphy]{alemi2018fixing}
Alemi, A., Poole, B., Fischer, I., Dillon, J., Saurous, R.~A., and Murphy, K.
\newblock Fixing a broken elbo.
\newblock In \emph{International Conference on Machine Learning}, 2018.

\bibitem[Ball{\'e} et~al.(2017)Ball{\'e}, Laparra, and
  Simoncelli]{balle2017end}
Ball{\'e}, J., Laparra, V., and Simoncelli, E.~P.
\newblock End-to-end optimized image compression.
\newblock \emph{International Conference on Learning Representations}, 2017.

\bibitem[Ball{\'e} et~al.(2018)Ball{\'e}, Minnen, Singh, Hwang, and
  Johnston]{balle2018variational}
Ball{\'e}, J., Minnen, D., Singh, S., Hwang, S.~J., and Johnston, N.
\newblock Variational image compression with a scale hyperprior.
\newblock \emph{In ICLR}, 2018.

\bibitem[Bamler \& Mandt(2017)Bamler and Mandt]{bamler2017dynamic}
Bamler, R. and Mandt, S.
\newblock Dynamic word embeddings.
\newblock In \emph{Proceedings of the 34th International Conference on Machine
  Learning-Volume 70}, pp.\  380--389, 2017.

\bibitem[Barkan(2017)]{barkan2017bayesian}
Barkan, O.
\newblock Bayesian neural word embedding.
\newblock In \emph{Association for the Advancement of Artificial Intelligence},
  pp.\  3135--3143, 2017.

\bibitem[Berger(1972)]{Berger1972OptimumQA}
Berger, T.
\newblock Optimum quantizers and permutation codes.
\newblock \emph{IEEE Trans. Information Theory}, 18:\penalty0 759--765, 1972.

\bibitem[Blei et~al.(2003)Blei, Ng, and Jordan]{blei2003latent}
Blei, D.~M., Ng, A.~Y., and Jordan, M.~I.
\newblock Latent dirichlet allocation.
\newblock \emph{Journal of machine Learning research}, 3:\penalty0 993--1022,
  2003.

\bibitem[Blei et~al.(2017)Blei, Kucukelbir, and McAuliffe]{blei2017variational}
Blei, D.~M., Kucukelbir, A., and McAuliffe, J.~D.
\newblock Variational inference: A review for statisticians.
\newblock \emph{Journal of the American statistical Association}, 112\penalty0
  (518):\penalty0 859--877, 2017.

\bibitem[Blundell et~al.(2015)Blundell, Cornebise, Kavukcuoglu, and
  Wierstra]{blundell2015weight}
Blundell, C., Cornebise, J., Kavukcuoglu, K., and Wierstra, D.
\newblock Weight uncertainty in neural networks.
\newblock \emph{arXiv preprint arXiv:1505.05424}, 2015.

\bibitem[Chou et~al.(1989)Chou, Lookabaugh, and
  Gray]{Chou1989EntropyconstrainedVQ}
Chou, P.~A., Lookabaugh, T.~D., and Gray, R.~M.
\newblock Entropy-constrained vector quantization.
\newblock \emph{IEEE Trans. Acoustics, Speech, and Signal Processing},
  37:\penalty0 31--42, 1989.

\bibitem[Cover \& Thomas(2012)Cover and Thomas]{cover2012elements}
Cover, T. and Thomas, J.
\newblock \emph{Elements of Information Theory}.
\newblock Wiley, 2012.
\newblock ISBN 9781118585771.
\newblock URL \url{https://books.google.com/books?id=VWq5GG6ycxMC}.

\bibitem[Gallager(1968)]{gallager1968information}
Gallager, R.~G.
\newblock \emph{Information theory and reliable communication}, volume~2.
\newblock Springer, 1968.

\bibitem[Gersho \& Gray(2012)Gersho and Gray]{gersho2012vector}
Gersho, A. and Gray, R.~M.
\newblock \emph{Vector quantization and signal compression}, volume 159.
\newblock Springer Science \& Business Media, 2012.

\bibitem[Gregor et~al.(2016)Gregor, Besse, Rezende, Danihelka, and
  Wierstra]{gregor2016conceptual}
Gregor, K., Besse, F., Rezende, D.~J., Danihelka, I., and Wierstra, D.
\newblock Towards conceptual compression, 2016.

\bibitem[Habibian et~al.(2019)Habibian, Rozendaal, Tomczak, and
  Cohen]{habibian2019video}
Habibian, A., Rozendaal, T.~v., Tomczak, J.~M., and Cohen, T.~S.
\newblock Video compression with rate-distortion autoencoders.
\newblock In \emph{Proceedings of the IEEE International Conference on Computer
  Vision}, pp.\  7033--7042, 2019.

\bibitem[Higgins et~al.(2017)Higgins, Matthey, Pal, Burgess, Glorot, Botvinick,
  Mohamed, and Lerchner]{higgins2016beta}
Higgins, I., Matthey, L., Pal, A., Burgess, C., Glorot, X., Botvinick, M.,
  Mohamed, S., and Lerchner, A.
\newblock beta-vae: Learning basic visual concepts with a constrained
  variational framework.
\newblock \emph{International Conference on Learning Representations}, 2017.

\bibitem[Honkela \& Valpola(2004)Honkela and Valpola]{honkela2004variational}
Honkela, A. and Valpola, H.
\newblock Variational learning and bits-back coding: an information-theoretic
  view to bayesian learning.
\newblock \emph{IEEE transactions on Neural Networks}, 15\penalty0
  (4):\penalty0 800--810, 2004.

\bibitem[Huffman(1952)]{huffman1952method}
Huffman, D.~A.
\newblock A method for the construction of minimum-redundancy codes.
\newblock \emph{Proceedings of the IRE}, 40\penalty0 (9):\penalty0 1098--1101,
  1952.

\bibitem[Johnston et~al.(2018)Johnston, Vincent, Minnen, Covell, Singh, Chinen,
  Jin~Hwang, Shor, and Toderici]{johnston2018improved}
Johnston, N., Vincent, D., Minnen, D., Covell, M., Singh, S., Chinen, T.,
  Jin~Hwang, S., Shor, J., and Toderici, G.
\newblock Improved lossy image compression with priming and spatially adaptive
  bit rates for recurrent networks.
\newblock In \emph{Proceedings of the IEEE Conference on Computer Vision and
  Pattern Recognition}, pp.\  4385--4393, 2018.

\bibitem[Jordan et~al.(1999)Jordan, Ghahramani, Jaakkola, and
  Saul]{jordan1999introduction}
Jordan, M.~I., Ghahramani, Z., Jaakkola, T.~S., and Saul, L.~K.
\newblock An introduction to variational methods for graphical models.
\newblock \emph{Machine learning}, 37\penalty0 (2):\penalty0 183--233, 1999.

\bibitem[Kingma \& Welling(2014{\natexlab{a}})Kingma and Welling]{KW2014}
Kingma, D.~P. and Welling, M.
\newblock Auto-encoding variational {Bayes}.
\newblock In \emph{International Conference on Learning Representations},
  2014{\natexlab{a}}.

\bibitem[Kingma \& Welling(2014{\natexlab{b}})Kingma and
  Welling]{kingma2013auto}
Kingma, D.~P. and Welling, M.
\newblock Auto-encoding variational {Bayes}.
\newblock In \emph{International Conference on Learning Representations}, pp.\
  1--9, 2014{\natexlab{b}}.

\bibitem[Kodak()]{kodak-cd}
Kodak, E.
\newblock {The Kodak PhotoCD Dataset}.
\newblock \url{https://www.cns.nyu.edu/~lcv/iclr2017/}.
\newblock Accessed: 2020-01-09.

\bibitem[Lloyd(1982)]{lloyd1982least}
Lloyd, S.
\newblock Least squares quantization in pcm.
\newblock \emph{IEEE transactions on information theory}, 28\penalty0
  (2):\penalty0 129--137, 1982.

\bibitem[Lombardo et~al.(2019)Lombardo, Han, Schroers, and
  Mandt]{lombardo2019deep}
Lombardo, S., Han, J., Schroers, C., and Mandt, S.
\newblock Deep generative video compression.
\newblock In \emph{Advances in Neural Information Processing Systems}, pp.\
  9283--9294, 2019.

\bibitem[MacKay(1992)]{mackay1992practical}
MacKay, D.~J.
\newblock A practical bayesian framework for backpropagation networks.
\newblock \emph{Neural computation}, 4\penalty0 (3):\penalty0 448--472, 1992.

\bibitem[MacKay(2003)]{mackay2003information}
MacKay, D.~J.
\newblock \emph{Information theory, inference and learning algorithms}.
\newblock Cambridge University Press, 2003.

\bibitem[Mentzer et~al.(2018)Mentzer, Agustsson, Tschannen, Timofte, and
  Van~Gool]{mentzer2018conditional}
Mentzer, F., Agustsson, E., Tschannen, M., Timofte, R., and Van~Gool, L.
\newblock Conditional probability models for deep image compression.
\newblock In \emph{Proceedings of the IEEE Conference on Computer Vision and
  Pattern Recognition}, pp.\  4394--4402, 2018.

\bibitem[Michel et~al.(2011)Michel, Shen, Aiden, Veres, Gray, Pickett, Hoiberg,
  Clancy, Norvig, Orwant, et~al.]{michel2011quantitative}
Michel, J.-B., Shen, Y.~K., Aiden, A.~P., Veres, A., Gray, M.~K., Pickett,
  J.~P., Hoiberg, D., Clancy, D., Norvig, P., Orwant, J., et~al.
\newblock Quantitative analysis of culture using millions of digitized books.
\newblock \emph{science}, 331\penalty0 (6014):\penalty0 176--182, 2011.

\bibitem[Mikolov et~al.(2013{\natexlab{a}})Mikolov, Chen, Corrado, and
  Dean]{mikolov2013efficient}
Mikolov, T., Chen, K., Corrado, G., and Dean, J.
\newblock Efficient estimation of word representations in vector space.
\newblock \emph{arXiv preprint arXiv:1301.3781}, 2013{\natexlab{a}}.

\bibitem[Mikolov et~al.(2013{\natexlab{b}})Mikolov, Sutskever, Chen, Corrado,
  and Dean]{mikolov2013distributed}
Mikolov, T., Sutskever, I., Chen, K., Corrado, G.~S., and Dean, J.
\newblock Distributed representations of words and phrases and their
  compositionality.
\newblock In \emph{Advances in Neural Information Processing Systems}, pp.\
  3111--3119, 2013{\natexlab{b}}.

\bibitem[Mnih \& Salakhutdinov(2008)Mnih and
  Salakhutdinov]{mnih2008probabilistic}
Mnih, A. and Salakhutdinov, R.~R.
\newblock Probabilistic matrix factorization.
\newblock In \emph{Advances in neural information processing systems}, pp.\
  1257--1264, 2008.

\bibitem[Ranganath et~al.(2014)Ranganath, Gerrish, and
  Blei]{ranganath2014black}
Ranganath, R., Gerrish, S., and Blei, D.~M.
\newblock Black box variational inference.
\newblock In \emph{International Conference on Artificial Intelligence and
  Statistics}, pp.\  814--822, 2014.

\bibitem[Rezende et~al.(2014)Rezende, Mohamed, and
  Wierstra]{rezende2014stochastic}
Rezende, D.~J., Mohamed, S., and Wierstra, D.
\newblock Stochastic backpropagation and approximate inference in deep
  generative models.
\newblock In \emph{International Conference on Machine Learning}, pp.\
  1278--1286, 2014.

\bibitem[Rippel \& Bourdev(2017)Rippel and Bourdev]{rippel2017realtime}
Rippel, O. and Bourdev, L.
\newblock Real-time adaptive image compression, 2017.

\bibitem[Theis et~al.(2017)Theis, Shi, Cunningham, and
  Husz{\'a}r]{theis2017lossy}
Theis, L., Shi, W., Cunningham, A., and Husz{\'a}r, F.
\newblock Lossy image compression with compressive autoencoders.
\newblock \emph{International Conference on Learning Representations}, 2017.

\bibitem[Toderici et~al.(2016)Toderici, O'Malley, Hwang, Vincent, Minnen,
  Baluja, Covell, and Sukthankar]{toderici2015variable}
Toderici, G., O'Malley, S.~M., Hwang, S.~J., Vincent, D., Minnen, D., Baluja,
  S., Covell, M., and Sukthankar, R.
\newblock Variable rate image compression with recurrent neural networks.
\newblock \emph{International Conference on Learning Representations}, 2016.

\bibitem[Toderici et~al.(2017)Toderici, Vincent, Johnston, Hwang, Minnen, Shor,
  and Covell]{toderici2017full}
Toderici, G., Vincent, D., Johnston, N., Hwang, S.~J., Minnen, D., Shor, J.,
  and Covell, M.
\newblock Full resolution image compression with recurrent neural networks.
\newblock In \emph{2017 IEEE Conference on Computer Vision and Pattern
  Recognition (CVPR)}, pp.\  5435--5443, 2017.

\bibitem[Townsend et~al.(2019)Townsend, Bird, and
  Barber]{townsend2019practical}
Townsend, J., Bird, T., and Barber, D.
\newblock Practical lossless compression with latent variables using bits back
  coding.
\newblock \emph{arXiv preprint arXiv:1901.04866}, 2019.

\bibitem[Wang et~al.(2003)Wang, Simoncelli, and Bovik]{wang2003multiscale}
Wang, Z., Simoncelli, E.~P., and Bovik, A.~C.
\newblock Multiscale structural similarity for image quality assessment.
\newblock In \emph{The Thrity-Seventh Asilomar Conference on Signals, Systems
  \& Computers, 2003}, volume~2, pp.\  1398--1402. Ieee, 2003.

\bibitem[Witten et~al.(1987)Witten, Neal, and Cleary]{witten1987arithmetic}
Witten, I.~H., Neal, R.~M., and Cleary, J.~G.
\newblock Arithmetic coding for data compression.
\newblock \emph{Communications of the ACM}, 30\penalty0 (6):\penalty0 520--540,
  1987.

\bibitem[Yang et~al.(2020)Yang, Bamler, and Mandt]{yang2020improving}
Yang, Y., Bamler, R., and Mandt, S.
\newblock Improving inference for neural image compression.
\newblock \emph{arXiv preprint arXiv:2006.04240}, 2020.

\bibitem[Zhang et~al.(2019)Zhang, Butepage, Kjellstrom, and
  Mandt]{zhang2019advances}
Zhang, C., Butepage, J., Kjellstrom, H., and Mandt, S.
\newblock Advances in variational inference.
\newblock \emph{IEEE transactions on pattern analysis and machine
  intelligence}, 2019.

\end{thebibliography}
\bibliographystyle{icml2020}

\end{document}


\renewcommand{\thesection}{S\arabic{section}}
\renewcommand{\theequation}{S\arabic{equation}}
\renewcommand{\thepage}{S\arabic{page}}
\renewcommand{\thetable}{S\arabic{table}}
\renewcommand{\thefigure}{S\arabic{figure}}

\twocolumn[
\icmltitle{Supplementary Material to \\ ``Variational Bayesian Quantization''}

\icmlsetsymbol{equal}{*}

\begin{icmlauthorlist}
\icmlauthor{Yibo Yang}{equal,to}
\icmlauthor{Robert Bamler}{equal,to}
\icmlauthor{Stephan Mandt}{to}
\end{icmlauthorlist}

\icmlaffiliation{to}{Department of Computer Science, University of California, Irvine}

\icmlcorrespondingauthor{Yibo Yang}{yibo.yang@uci.edu}
\icmlcorrespondingauthor{Robert Bamler}{rbamler@uci.edu}

\icmlkeywords{Machine Learning, ICML}

\vskip 0.3in
]

\printAffiliationsAndNotice{\icmlEqualContribution} %

This document provides details of the proposed compression method (Section~\ref{app:practical-details}), model and experiment details (Section~\ref{app:exp-details}), and additional examples of compressed images (Section~\ref{app:examples}).

\section{Delimitation Overhead}
\label{app:practical-details}

We elaborate on the ``encoding'' paragraph of Section~3.3 of the main paper.
After finding a quantized code point~$\hat\xi_i$ for each dimension $i\in\{1,\ldots,K\}$ of the latent space, these code points have to be losslessly encoded into a single bitstring for transmission or storage.
We experimented with two encoding schemes, described in Subsections~\ref{app:alternative-encoding} and~\ref{app:used-encoding} below.
Subsection~\ref{app:rate-vs-inf-content} provides further analysis.

\subsection{Encoding via Concatenation}
\label{app:alternative-encoding}

We first describe an encoding scheme that we did not end up using, but that makes it easier to understand the objective function of VBQ (Eq.~8 of the main text).
This encoding scheme concatenates the binary representations of $\hat\xi_i$ (Eq.~4 of the main text) for all $i\in\{1,\ldots,K\}$ in to a single bitstring.
As each dimension~$i$ contributes $\rate(\hat\xi_i)$ bits to the concatenated bitstring, this encoding scheme justifies the rate penalty term ``$\lambda\rate(\hat\xi_i)$'' in Eq.~4 of the main text.

One also has to transmit the rates $\rate(\hat\xi_i)$ (in compressed form using traditional entropy coding) so that the decoder can split the concatenated bitstring at the correct positions.
While this incurs some overhead, the variable-bitlength representation of $\hat\xi_i$ also saves one bit per dimension~$i$ because the last bit in the binary representation of each~$\hat\xi_i$ does not need to be transmitted as it is always equal to one (otherwise, the optimization algorithm in VBQ would favor an equivalent shorter binary representation of~$\hat\xi_i$).

\subsection{Encoding via Standard Entropy Coding}
\label{app:used-encoding}

The actual encoding scheme we ended up using does not deal with the binary representation of each~$\hat\xi_i$ explicitly.
Instead, we treat each~$\hat\xi_i$ as a discrete symbol and directly encode the sequence $(\hat\xi_i)_{i=1}^K$ of symbols via entropy coding (e.g., standard arithmetic coding).
The entropy coder needs a model of the probability $p(\hat\xi_i)$ of each symbol.
For model compression, we use the empirical frequencies, which we transmit as extra header information that counts towards the total bitrate.
For data compression, we estimate the frequencies on training data and include them in the decoder.

Transmitting the empirical frequencies lead to a negligible overhead in the word embeddings experiment.
Only a few hundred code points (depending on~$\lambda$) had nonzero frequencies, so that the compressed file size was dominated by the encoding of $K = Vd = 10^7$ quantized latent variables.

\subsection{Justification of the Rate Penalty Term $\lambda\rate(\hat\xi_i)$}
\label{app:rate-vs-inf-content}

All experimental results are reported with the encoding scheme of Section~\ref{app:used-encoding} as it lead to slightly lower bitrates in practice.
A peculiarity of this encoding scheme is that it ignores the length $\rate(\hat\xi_i)$ of the binary representation of each~$\hat\xi_i$.
For a sequence of symbols $\boldsymbol{\hat\xi}\equiv (\hat\xi_i)_{i=1}^K$ with an i.i.d.~entropy model $p(\hat\xi_i)$, an optimal entropy coder (such as arithmetic coding) achieves the total bitrate $\rate(\boldsymbol{\hat\xi}) = \lceil h(\boldsymbol{\hat\xi}) \rceil$ with the information content
\begin{align}\label{app-eq:total-inf-content}
    h(\boldsymbol{\hat\xi})
    &= \sum_{i=1}^K h(\hat\xi_i)
    = -\sum_{i=1}^K \log_2 p(\hat\xi_i).
\end{align}

In particular, Eq.~\ref{app-eq:total-inf-content} does not depend on $\rate(\hat\xi_i)$.
This poses the question whether the rate penalty term ``$\lambda \rate(\hat\xi_i)$'' in the VBQ objective (Eq.~8 of the main text) is justified.
Ideally, the algorithm would minimize $\lambda h(\hat\xi_i)=-\lambda \log_2 p(\hat\xi_i)$ instead, but this quantity is unknown until the quantizations~$(\hat\xi_i)_{i=1}^K$ and therefore the empirical frequencies $p(\hat\xi_i)$ are obtained.
Our experiments suggest that $\rate(\hat\xi_i)$ is a useful proxy for the eventual value of $h(\hat\xi_i)$.

\begin{figure}[t]
\begin{center}
\includegraphics[width=\columnwidth]{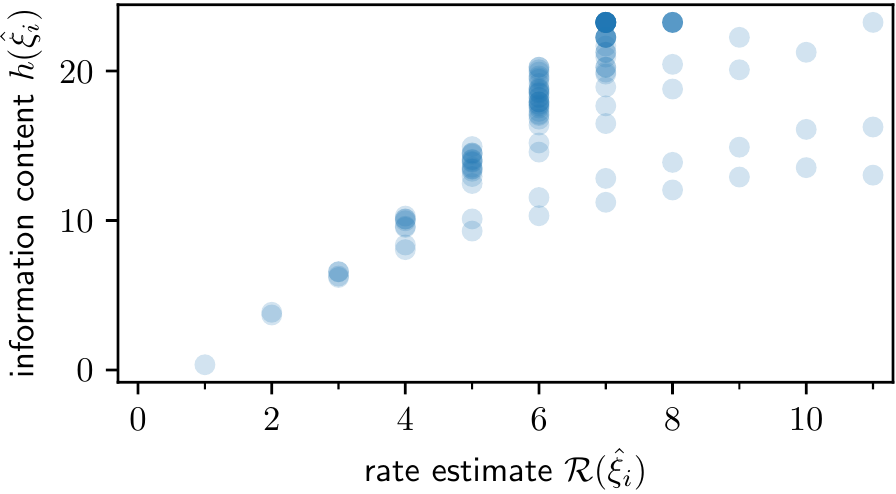}
\end{center}
\caption{Relation between rate estimate $\rate(\hat\xi_i)$ and the actual contribution $h(\hat\xi_i)$ of code point~$\hat\xi_i$ to the total bitrate under entropy coding.
The approximate affine linear relationship justifies minimizing $\rate(\hat\xi_i)$ as a proxy for $h(\hat\xi_i)$ in VBQ.}
\label{app-fig:rate-scatter}
\end{figure}

Figure~\ref{app-fig:rate-scatter} plots the rate estimate $\rate(\hat\xi_i)$, i.e., the integer number of bits in the binary representation of~$\hat\xi_i$ ($x$-axis) against the actual contribution $h(\hat\xi_i)=-\log_2 p(\hat\xi_i)$ to the total bitrate according to Eq.~\ref{app-eq:total-inf-content} ($y$-axis).
The figure shows experimental data for compressed word embeddings at $1.32$~bits per latent dimension.
We make the following observations:
\begin{itemize}
    \item
        For most code points~$\hat\xi_i$, the dependency between $h(\hat\xi_i)$ and $\rate(\hat\xi_i)$ can be approximated by an affine linear function, thus justifying the use of $\rate(\hat\xi_i)$ in the optimization of Bayesian~AC.
    \item
        The slope of the approximate linear dependency is larger than one.
        This may be understood by the penalty term $\lambda \rate(\hat\xi_i)$ in the objective function (Eq.~8 of the main text), which causes the method to avoid code points $\hat\xi_i$ with large rate estimates $\rate(\hat\xi_i)$, thus reducing their empirical frequencies $p(\hat\xi_i)$ and increasing their information content $h(\hat\xi_i)=-\log_2 p(\hat\xi_i)$.
        This observation does not invalidate the use of $\rate(\hat\xi_i)$ as an estimate for $h(\hat\xi_i)$ since the different slope can be absorbed in a rescaling of the parameter~$\lambda$ %
    \item
        For rates $\rate(\hat\xi_i)\geq 4$, there are two code points for each rate with considerably lower information content.
        These code points correspond to the two extremes for each rate, i.e., $\hat\xi_i$~closest to zero or one, respectively.
        The observation that the two extremes have lower information content (i.e., higher empirical frequencies) can be explained by the fact that the empirical prior distribution whose CDF we use to map latent variables~$z_i$ to quantiles~$\xi_i$ does not fully capture the true distribution of variational means.
        Indeed, experiments with a more long tailed empirical prior distribution lead to marginally better performance, but the simplicity of a Gaussian empirical prior seemed more valuable to us.
\end{itemize}

\section{More Experimental Details}
\label{app:exp-details}

\subsection{Word Embeddings}

The word embeddings experiment involved only minimal hyperparameter tuning, and we only optimized for performance of the uncompressed model since the goal of the experiment was to test the proposed compression method on a model that was not tuned for compression.
We trained for $10^5$ iterations with minibatches of $10^4$ randomly drawn words and contexts due to hardware constraints.
We tried learning rates $0.1$ and $1$ and chose $0.1$.

\subsection{Experiments on Images}

As mentioned in the main text, we used regular VAEs in the image experiments with standard normal prior $p(\bz)$ and factorized normal posterior $q(\bz|\bx)$ with diagonal covariance.

\subsection{MNIST}
The VAE's inference network has two convolutional layers followed by a fully connected layer. 
The two conv layers use 32 and 64 filters respectively, with kernel size 3, stride size 2, and ReLU activation. The fully connected layer has output dimension 10 so that $\boldsymbol\mu$ and $\boldsymbol\sigma^2$ of $q(\bz|\bx)$ each has dimension 5. 

The generative network architecture mirrors the inference network but in reverse, starting with a dense layer mapping 5 dimensional latent variables to 1568 dimensional, treated as 32-channel 7x7 activations,  and followed by two deconvolutional layers of 64 and 32 filters (with identical padding and stride as the convolutional layers). The output is deconvolved with a single 3x3 filter with sigmoid activation function.
For each pixel, the (scalar) output of the last layer parameterizes the likelihood of the pixel being white.

We trained the network on binarized MNIST images for 100 epochs, using the Adam optimizer with learning rate $10^{-4}$.

\subsection{Frey Faces}
On the Frey Faces dataset, we observe poor reconstruction quality by training on binarized images with a factorized Bernoulli likelihood model; instead, we treat each pixel as an observation from a factorized categorical likelihood model with 256 possible outcomes.

The VAE's inference network has two layers. The first layer flattens the input image, converts each pixel value in $\{0, 1, ..., 255\}$ into a one-hot vector $\in \mathbb{R}^{256}$, and uses it to index a 128-dimensional dense vector. The second layer flattens the result of the first layer as its input (which has dimensionality equal to 128 $\times$  \emph{number of pixels}), and fully connects its input to 8 hidden units. The final output is split to obtain 4-dimensional $\boldsymbol \mu$ and $\log \boldsymbol \sigma^2$ of $q(\bz|\bx)$.

The generative network has two fully connected layers. The first layer uses 4 hidden units and ReLU activation; the second layer uses   256 $\times$ \emph{number of pixels} hidden units, and takes a 256-way softmax to compute the categorical probability of of each pixel value taking value in $\{0, 1, ..., 255\}$.

We obtained the Frey Faces images from \url{https://cs.nyu.edu/~roweis/data.html}. We trained on a random subset of 1800 images for 800 epochs, using the Adam optimizer with learning rate $10^{-4}$.

On both MNIST and Frey Faces, we vary the rate-distortion trade-off parameter $\lambda$ of Variational Bayesian Quantization between $10^{-5}$ and $10^4$.

\subsection{ Color Image Compression}

As mentioned in the main text, the VAE here uses a fully convolutional architecture with 3 layers of 256 filters each, the same as in (Ball{\'e} et al., 2017); see the latter for detailed descriptions.
We tuned the variance $\sigma^2$ of the likelihood model on a logarithmic grid from $10^{-4}$ to $0.1$ and set it to $0.001$. The VAE was trained on the same dataset as in (Ball{\'e} et al., 2017) for 2 million steps, using Adam with learning rate $10^{-4}$.

In the image compression R-D curves, $\lambda$ ranges from $2^{-6}$ to $2^{16}$.
In Figure 5 of the main text, $\lambda$ was set to $17.5$ for VBQ to match the bitrate of the other methods. The uniform quantization result was obtained with 4 quantization levels, on a separately tuned model that had an additional convolutional layer of 64 channels. The additional conv layer was to reduce the latent dimensionality, as uniform quantization could not achieve bitrates lower than 0.5 even with only 2 grid points in the original 3-layer model.

\section{Additional Image Compression Examples}
\label{app:examples}

Starting on the next page, we provide detailed compression results for individual images from the Kodak dataset.
For each image, we show the rate-distortion performance by various methods, followed by reconstructions using our proposed method and JPEG at equal bitrate.%
\footnote{The present version of this document contains a subset of example images due to a file size limit on arXiv submissions.}

\newpage
\onecolumn

\begin{figure}[t]
\centering
 \begin{subfigure}[h]{0.45\columnwidth}
 \centering
 \includegraphics[width=\textwidth]{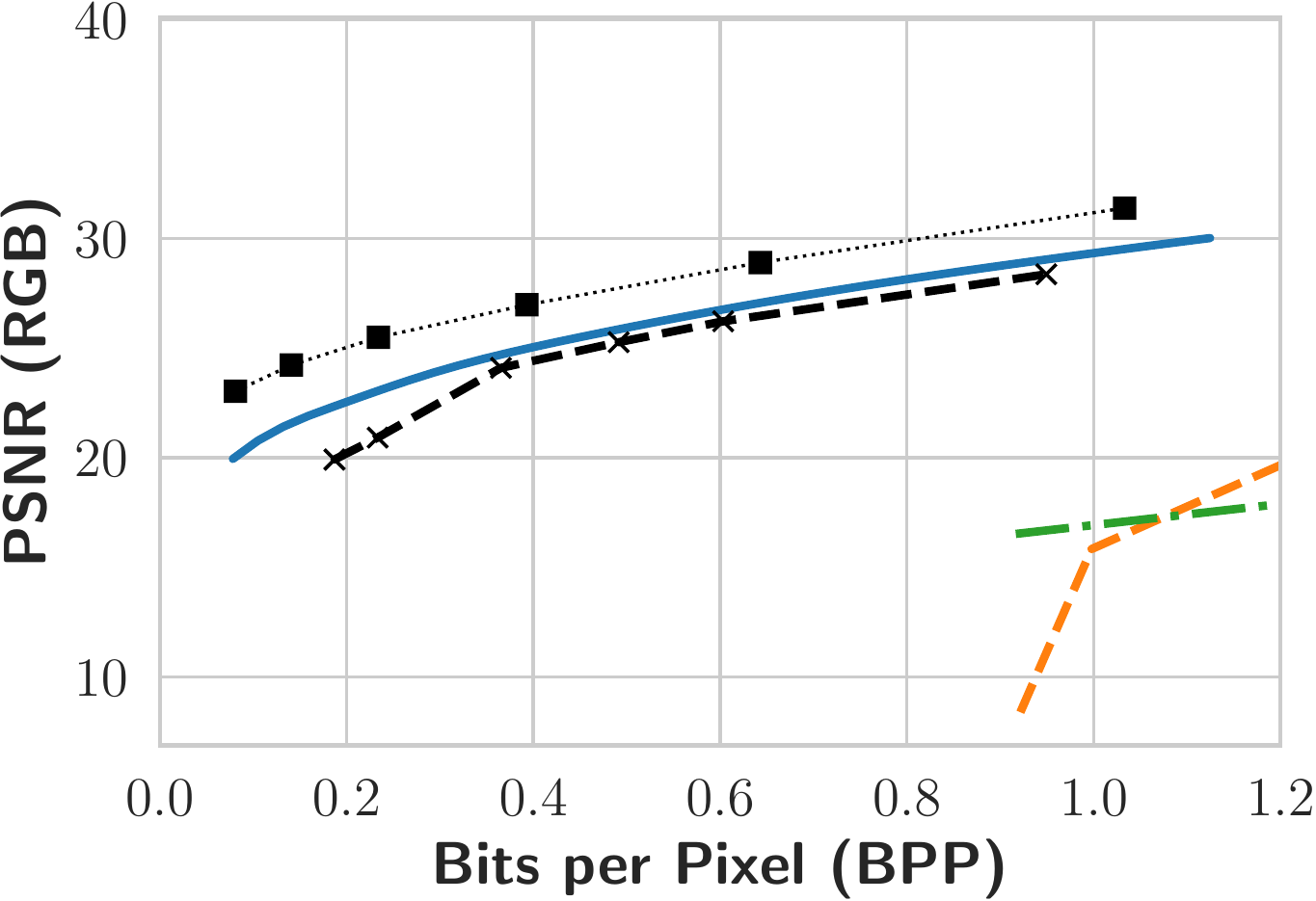}
 \end{subfigure}
 \hfill
 \begin{subfigure}[h]{0.45\columnwidth}
 \centering
 \includegraphics[width=\textwidth]{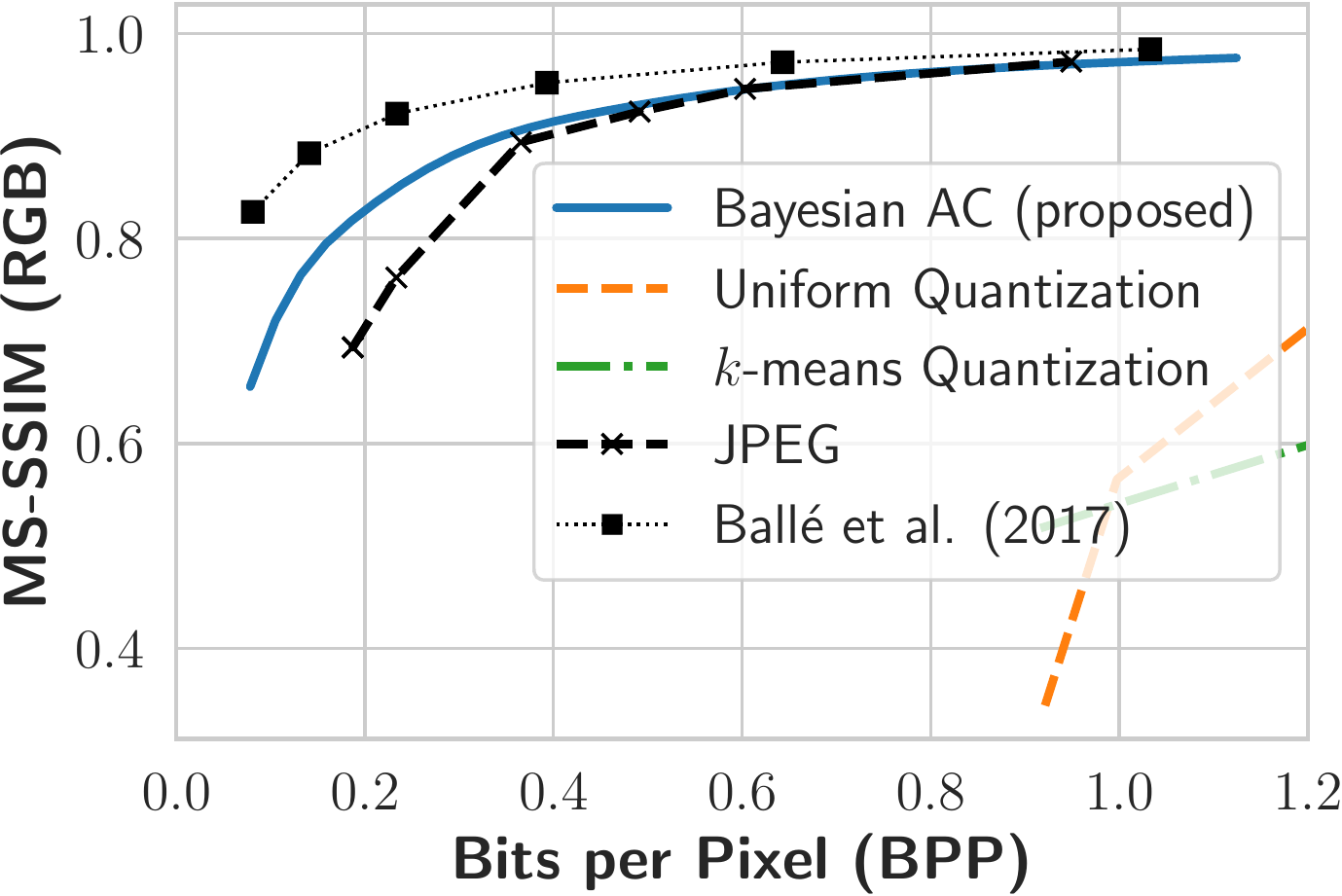}
 \end{subfigure}

 \vspace{10pt}\includegraphics[width=0.7\textwidth]{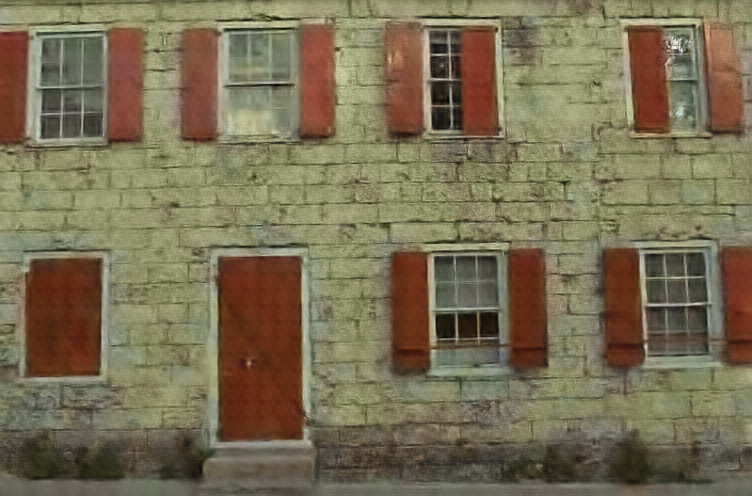}\vspace{-10pt}\caption{Proposed. bits-per-pixel: 0.27, PSNR: 23.595, MS-SSIM: 0.871}
 \vspace{10pt}\includegraphics[width=0.7\textwidth]{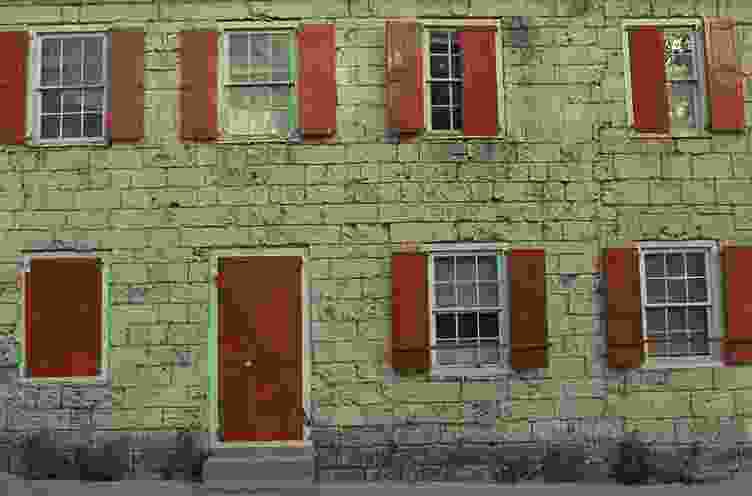}\vspace{-10pt}\caption{JPEG. bits-per-pixel: 0.27, PSNR: 22.015, MS-SSIM: 0.816}
\end{figure}

\begin{figure}[t]
\centering
 \begin{subfigure}[h]{0.45\columnwidth}
 \centering
 \includegraphics[width=\textwidth]{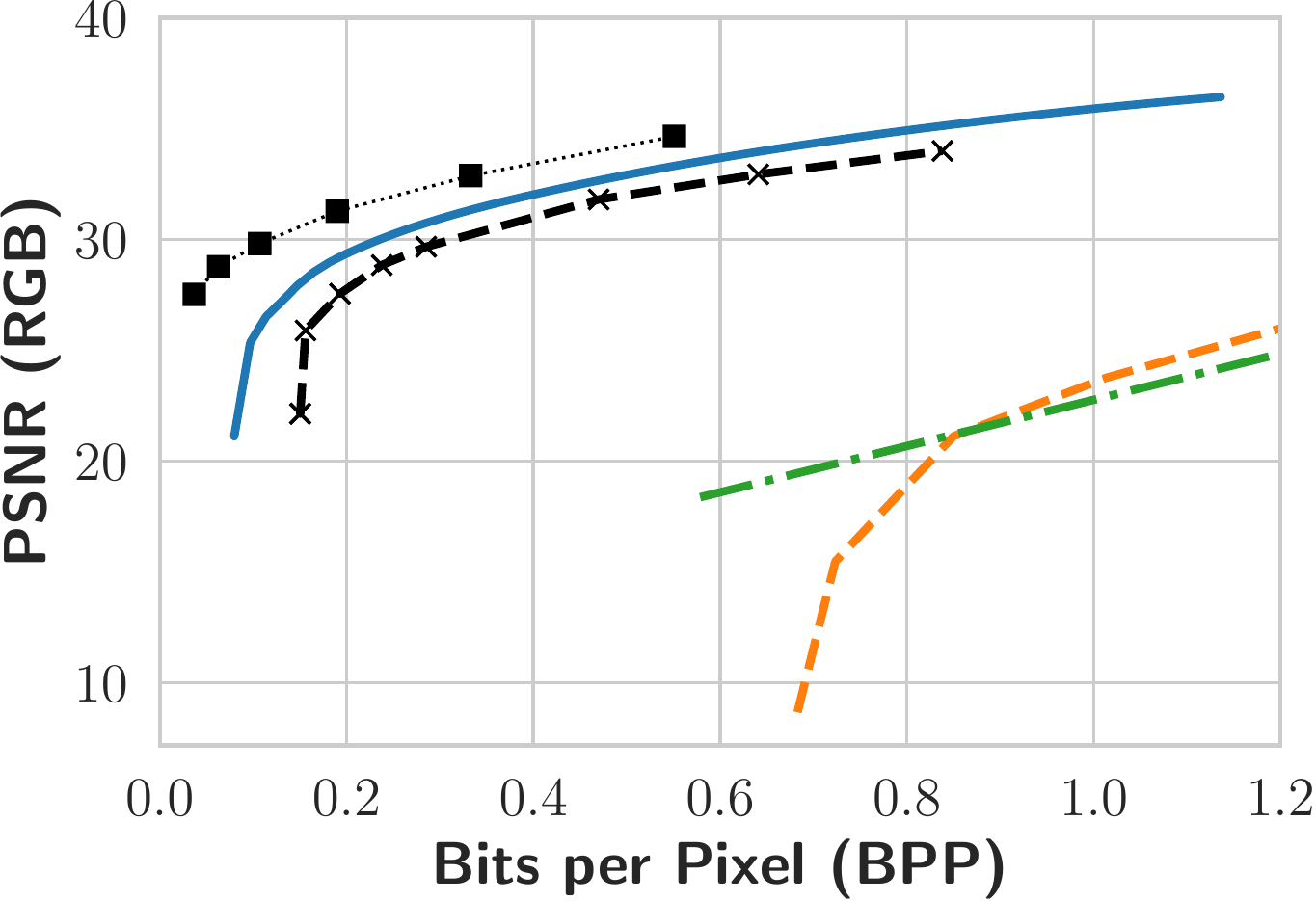}
 \end{subfigure}
 \hfill
 \begin{subfigure}[h]{0.45\columnwidth}
 \centering
 \includegraphics[width=\textwidth]{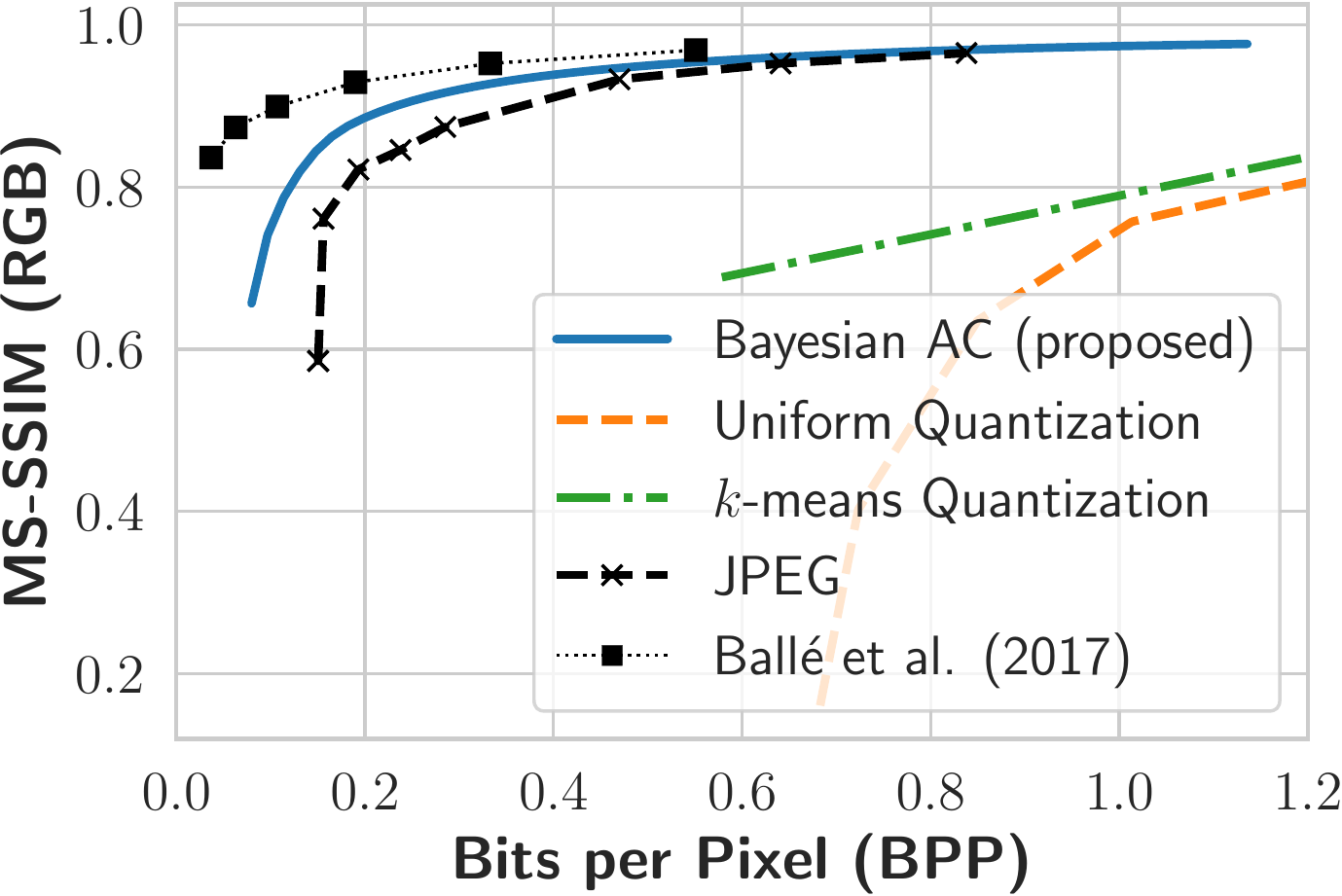}
 \end{subfigure}

 \vspace{10pt}\includegraphics[width=0.7\textwidth]{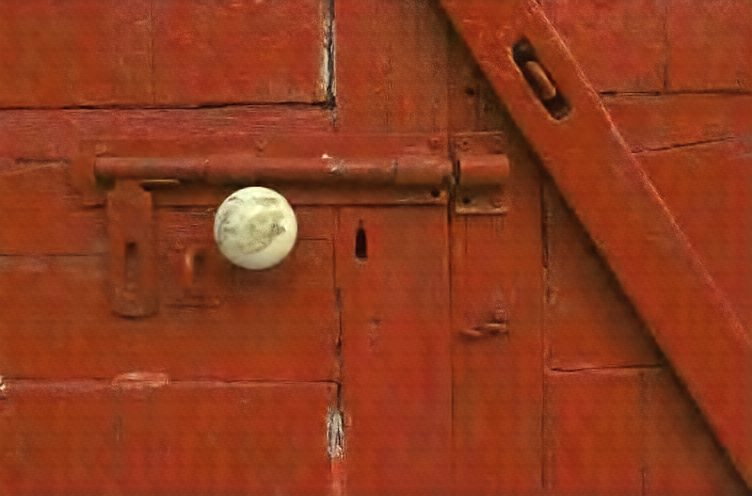}\vspace{-10pt}\caption{Proposed. bits-per-pixel: 0.19, PSNR: 29.226, MS-SSIM: 0.882}
 \vspace{10pt}\includegraphics[width=0.7\textwidth]{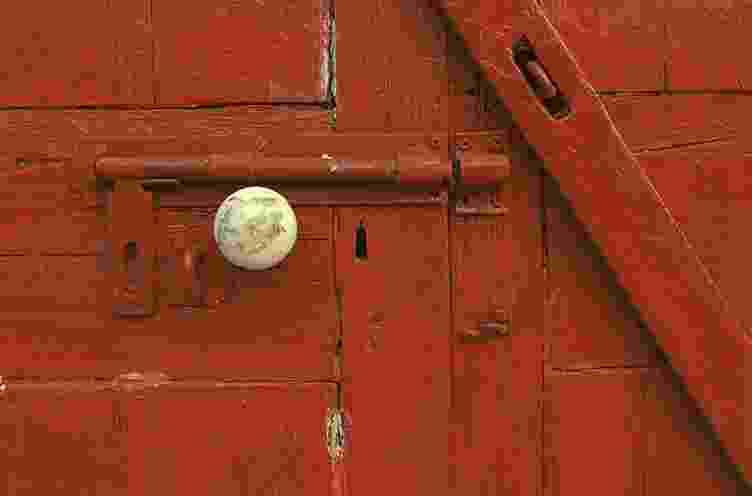}\vspace{-10pt}\caption{JPEG. bits-per-pixel: 0.19, PSNR: 26.658, MS-SSIM: 0.747}
\end{figure}

\begin{figure}[t]
\centering
 \begin{subfigure}[h]{0.45\columnwidth}
 \centering
 \includegraphics[width=\textwidth]{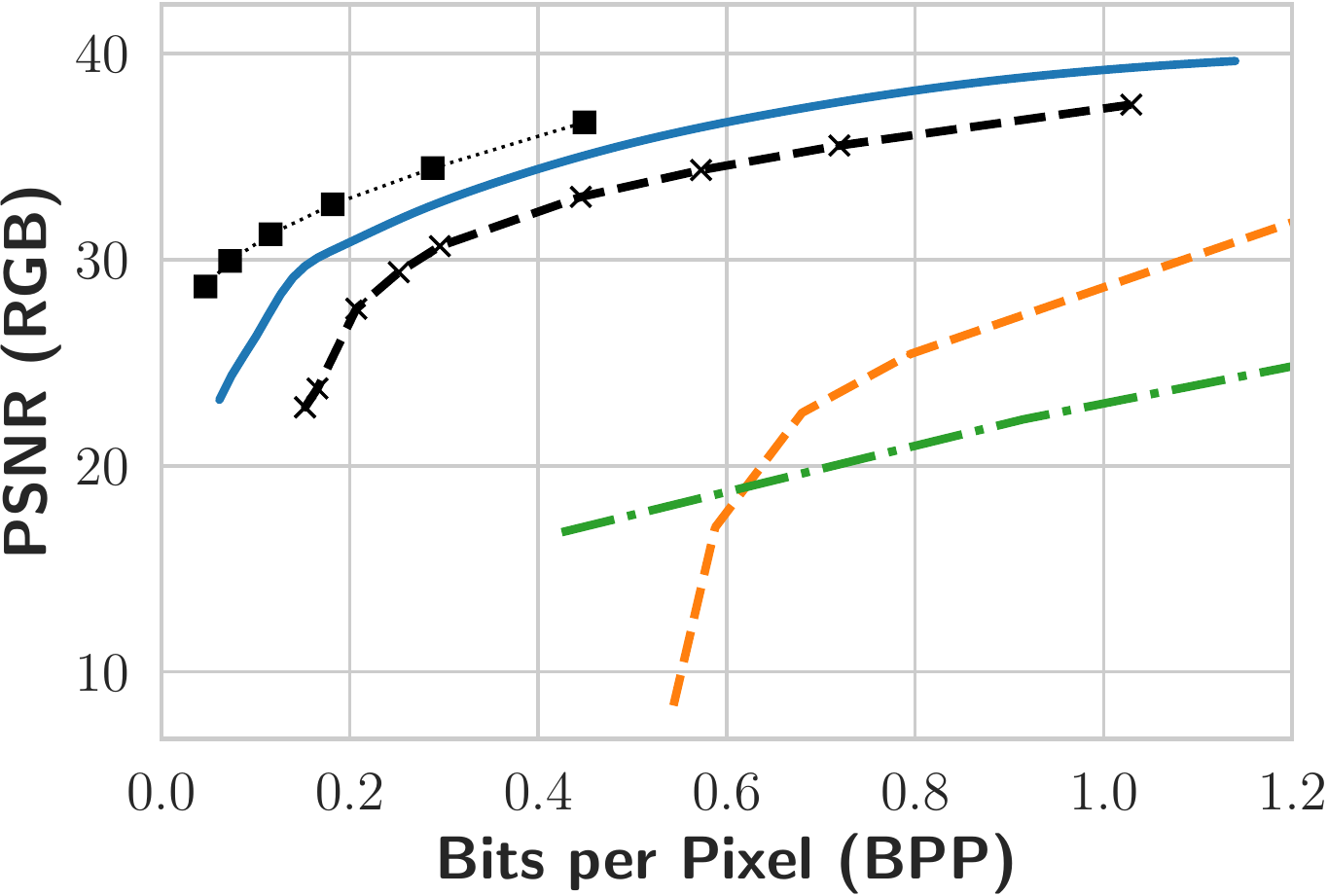}
 \end{subfigure}
 \hfill
 \begin{subfigure}[h]{0.45\columnwidth}
 \centering
 \includegraphics[width=\textwidth]{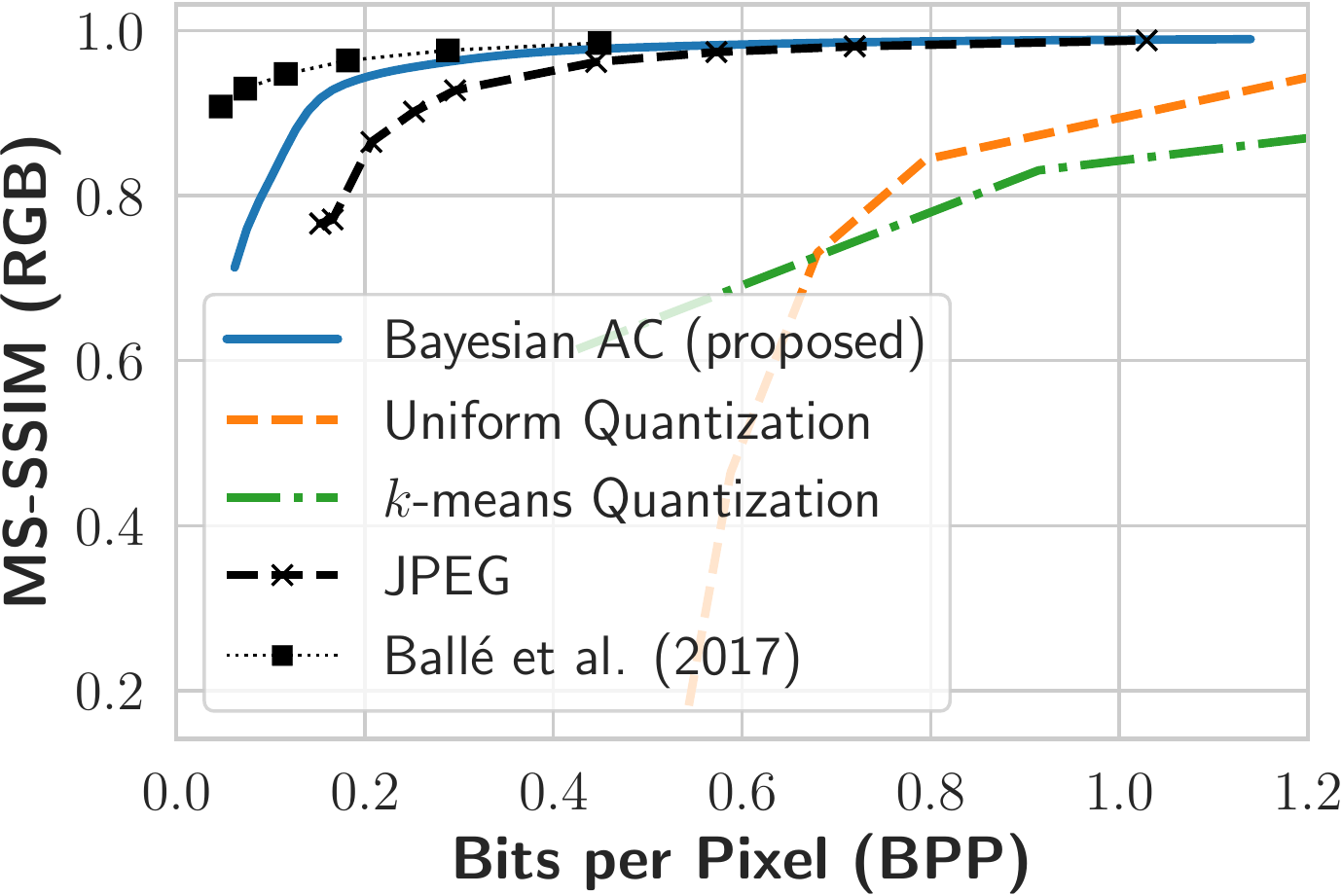}
 \end{subfigure}

 \vspace{10pt}\includegraphics[width=0.7\textwidth]{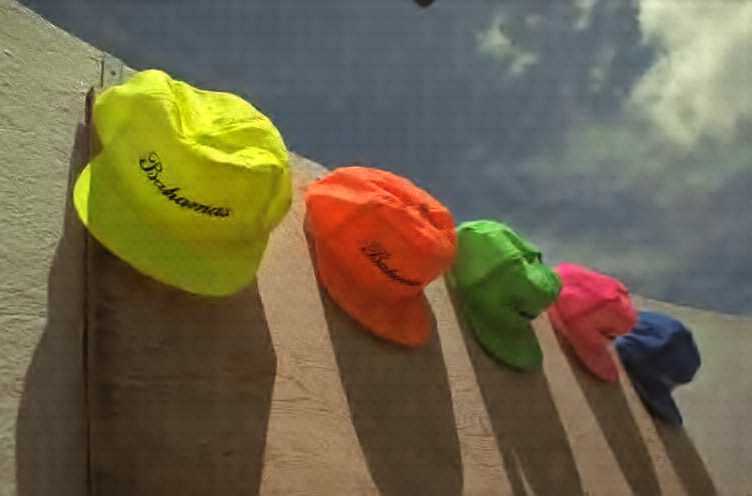}\vspace{-10pt}\caption{Proposed. bits-per-pixel: 0.19, PSNR: 30.664, MS-SSIM: 0.94}
 \vspace{10pt}\includegraphics[width=0.7\textwidth]{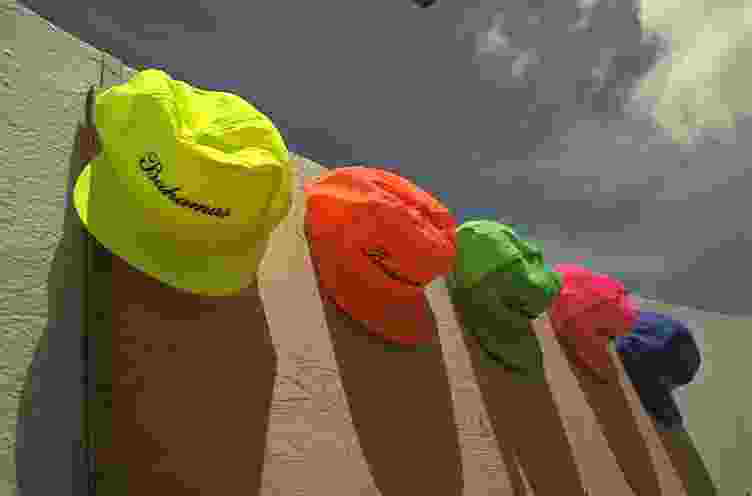}\vspace{-10pt}\caption{JPEG. bits-per-pixel: 0.19, PSNR: 26.201, MS-SSIM: 0.842}
\end{figure}

\begin{figure}[t]
\centering
 \begin{subfigure}[h]{0.45\columnwidth}
 \centering
 \includegraphics[width=\textwidth]{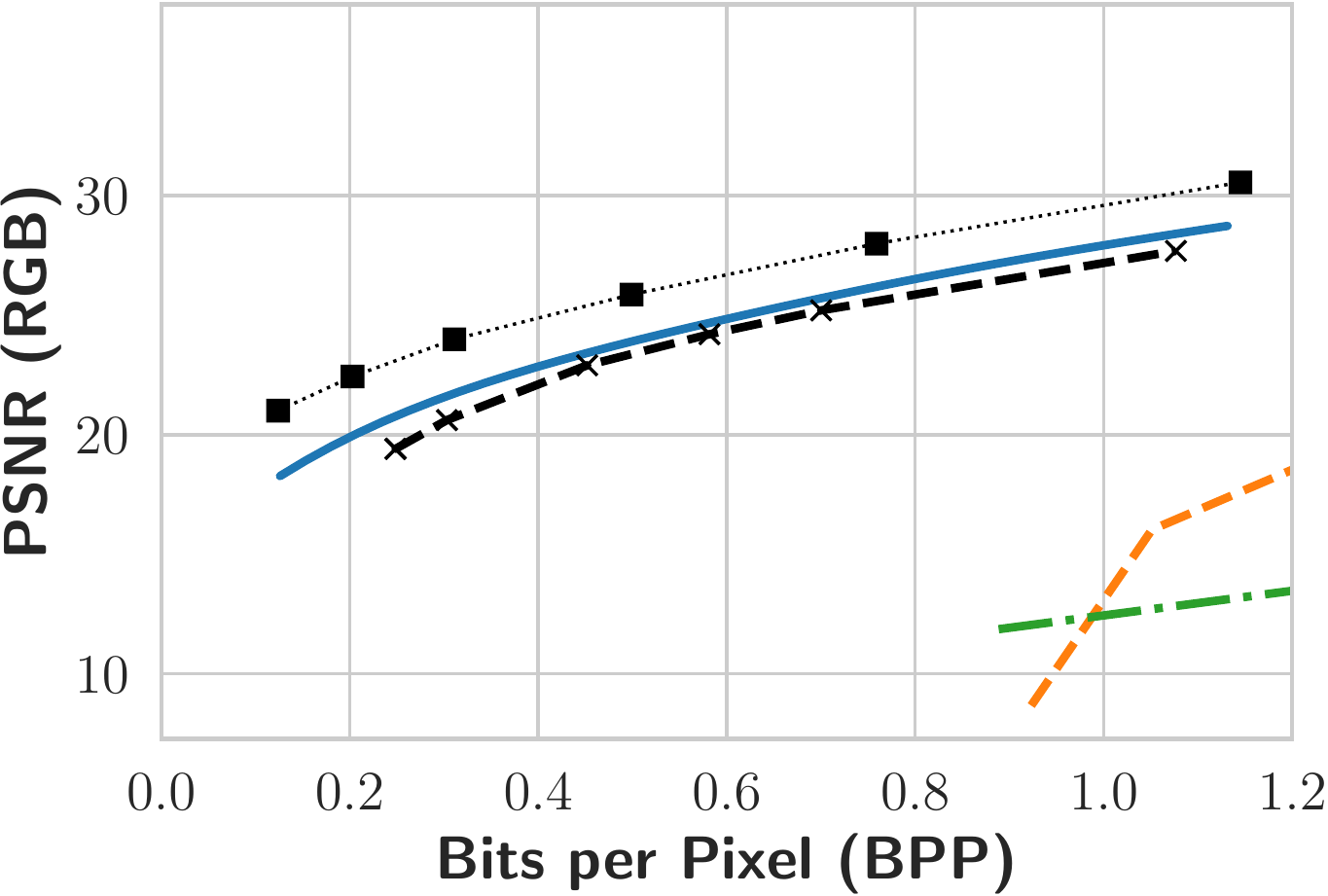}
 \end{subfigure}
 \hfill
 \begin{subfigure}[h]{0.45\columnwidth}
 \centering
 \includegraphics[width=\textwidth]{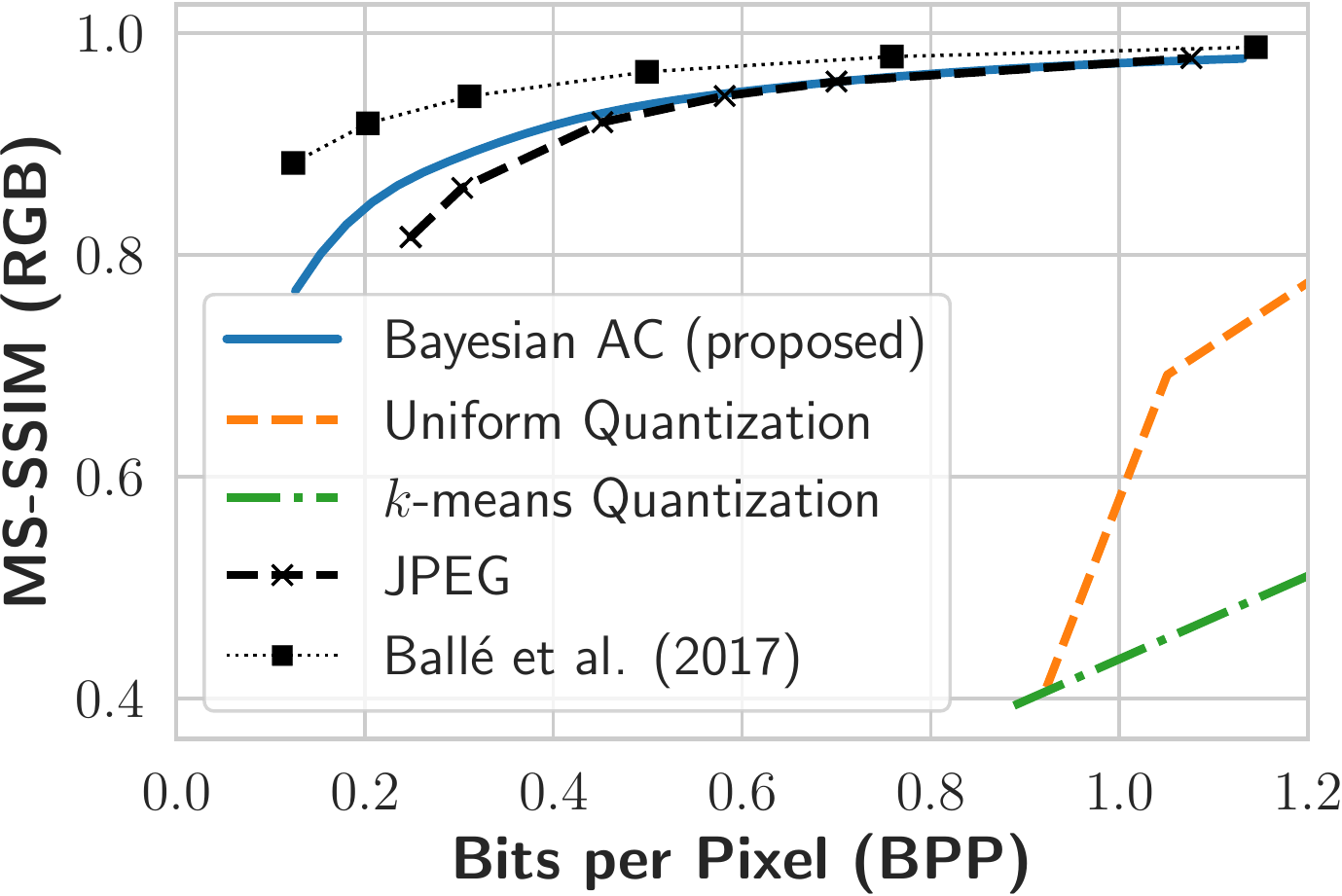}
 \end{subfigure}

 \vspace{10pt}\includegraphics[width=0.7\textwidth]{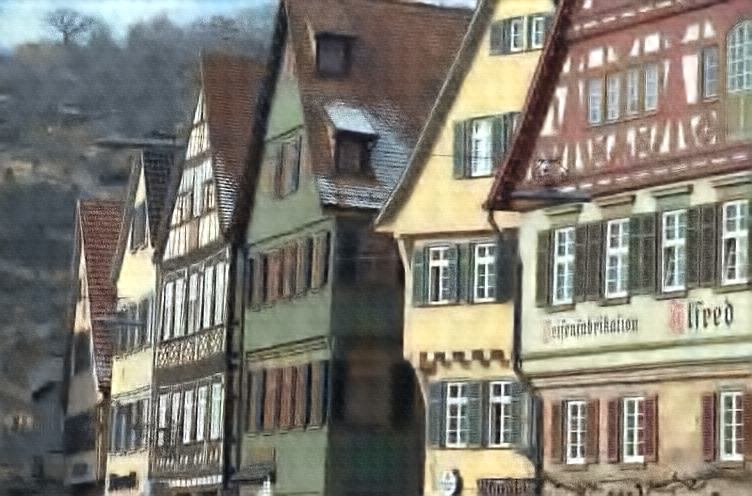}\vspace{-10pt}\caption{Proposed. bits-per-pixel: 0.34, PSNR: 22.177, MS-SSIM: 0.903}
 \vspace{10pt}\includegraphics[width=0.7\textwidth]{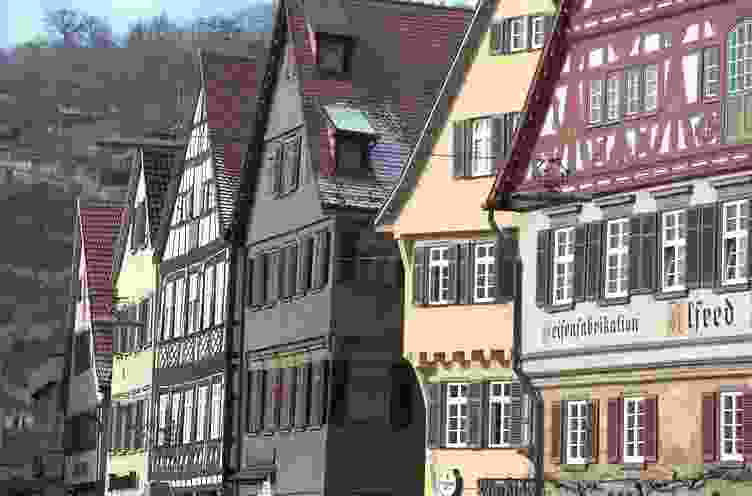}\vspace{-10pt}\caption{JPEG. bits-per-pixel: 0.34, PSNR: 21.331, MS-SSIM: 0.882}
\end{figure}

\begin{figure}[t]
\centering
 \begin{subfigure}[h]{0.45\columnwidth}
 \centering
 \includegraphics[width=\textwidth]{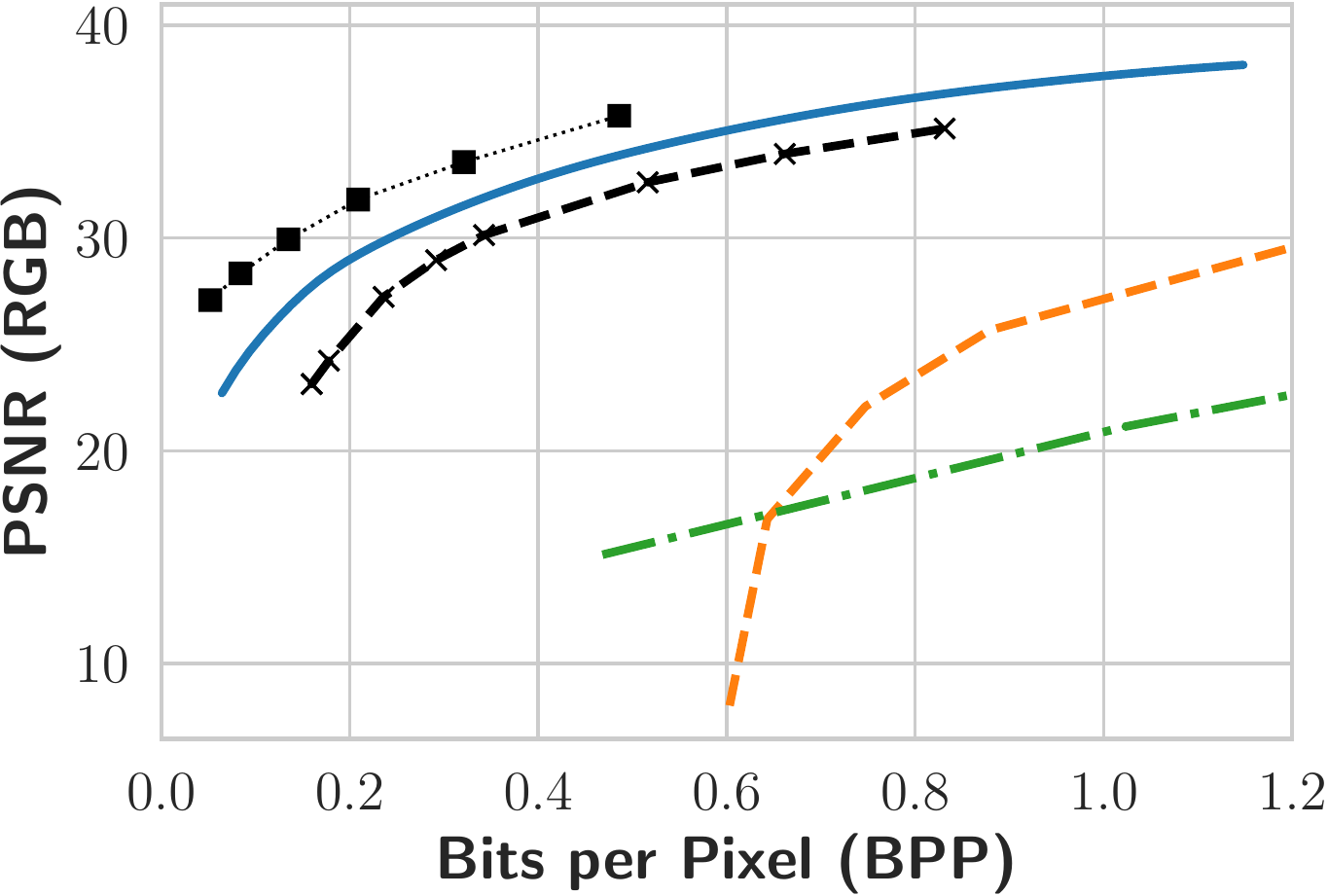}
 \end{subfigure}
 \hfill
 \begin{subfigure}[h]{0.45\columnwidth}
 \centering
 \includegraphics[width=\textwidth]{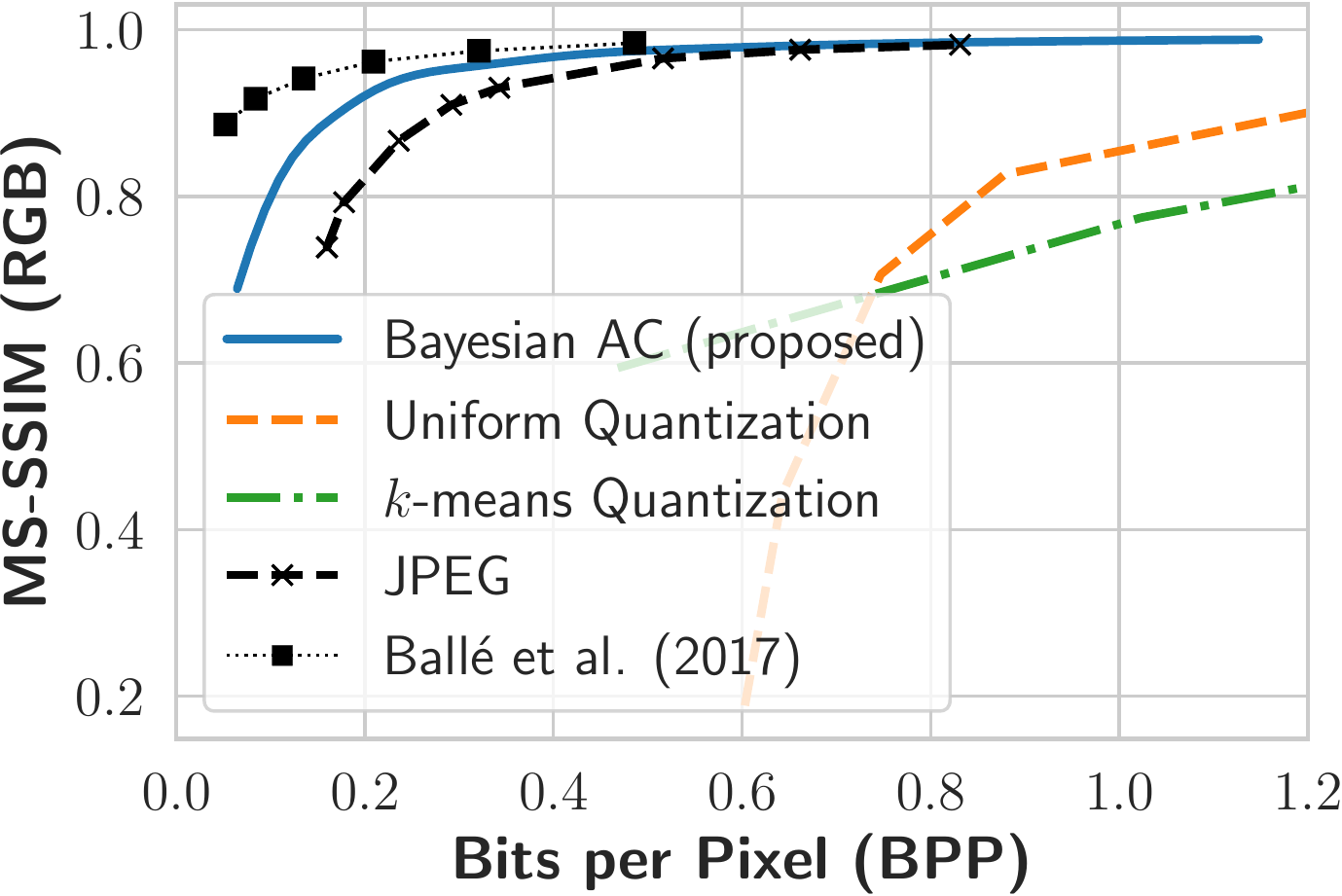}
 \end{subfigure}

 \vspace{10pt}
 \begin{subfigure}[h]{0.49\columnwidth}
 \centering
 \includegraphics[width=\textwidth]{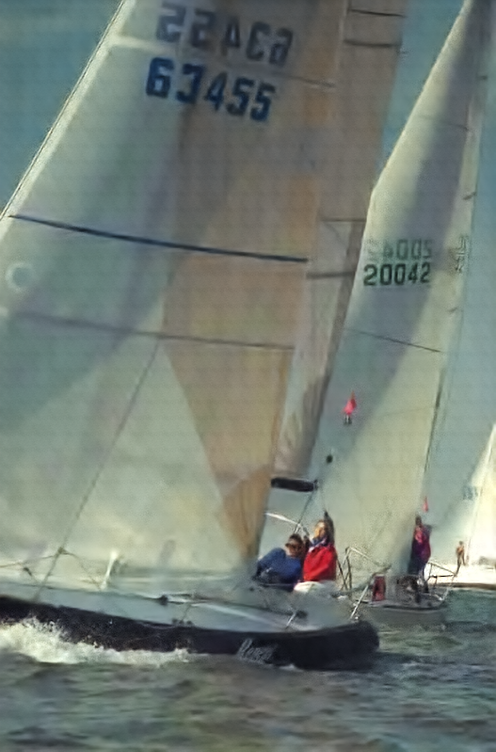}\caption{Proposed. bits-per-pixel: 0.22, PSNR: 29.584, MS-SSIM: 0.935}
 \end{subfigure}
 \hfill
 \begin{subfigure}[h]{0.49\columnwidth}
 \centering
 \includegraphics[width=\textwidth]{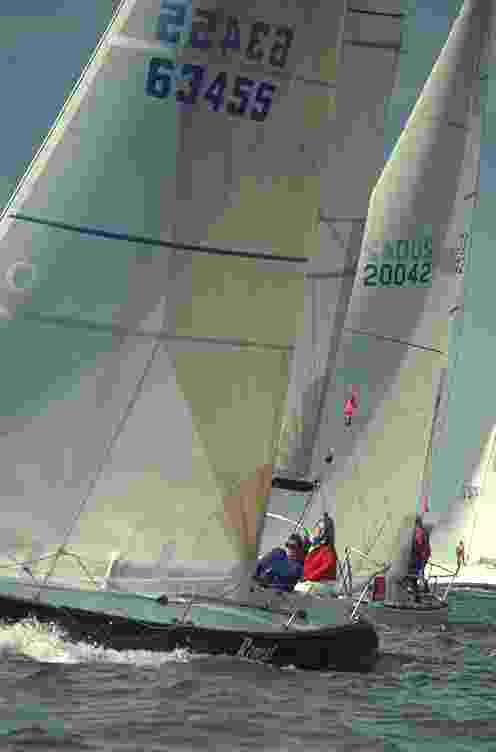}\caption{JPEG. bits-per-pixel: 0.22, PSNR: 26.543, MS-SSIM: 0.846}
 \end{subfigure}
\end{figure}

\begin{figure}[t]
\centering
 \begin{subfigure}[h]{0.45\columnwidth}
 \centering
 \includegraphics[width=\textwidth]{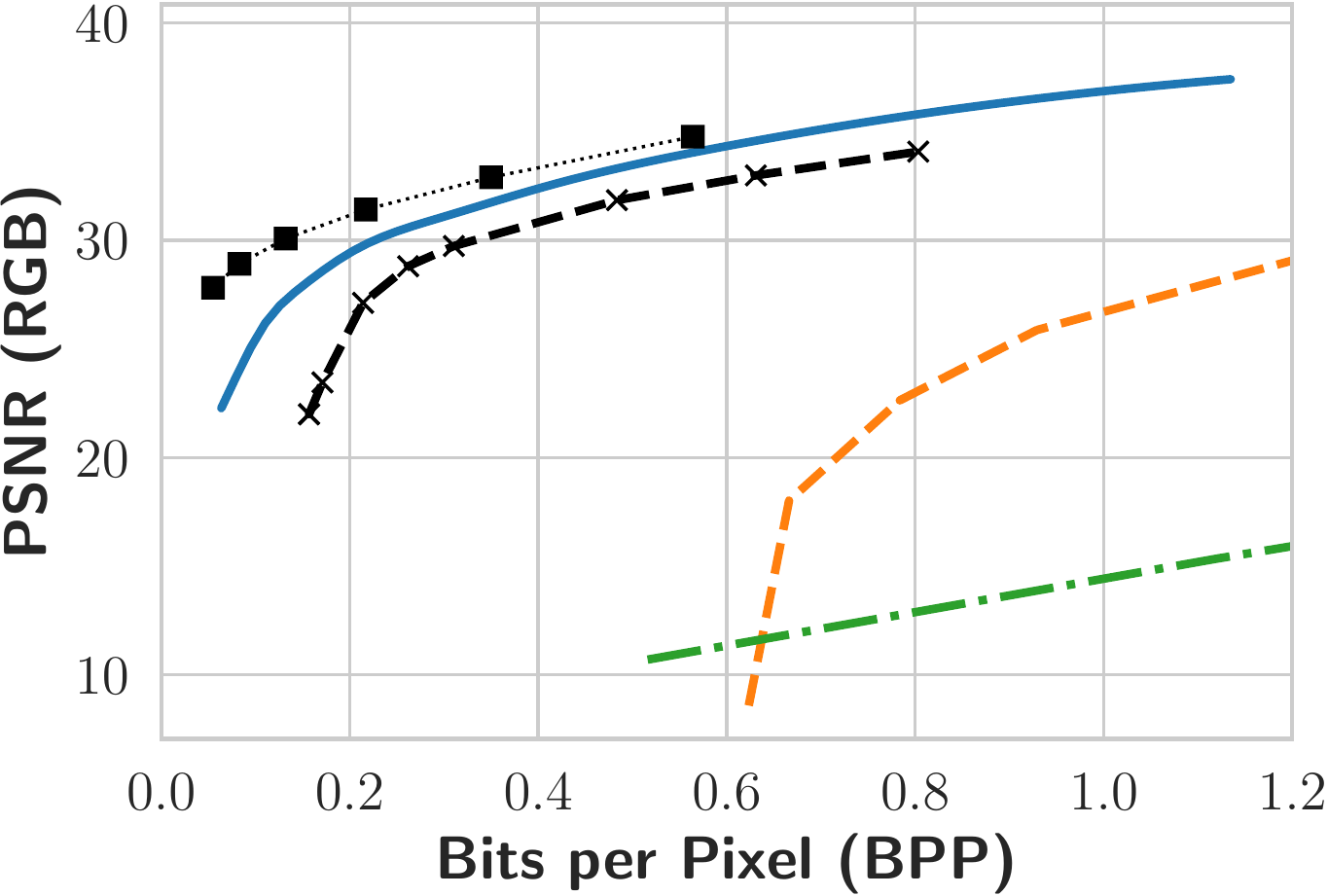}
 \end{subfigure}
 \hfill
 \begin{subfigure}[h]{0.45\columnwidth}
 \centering
 \includegraphics[width=\textwidth]{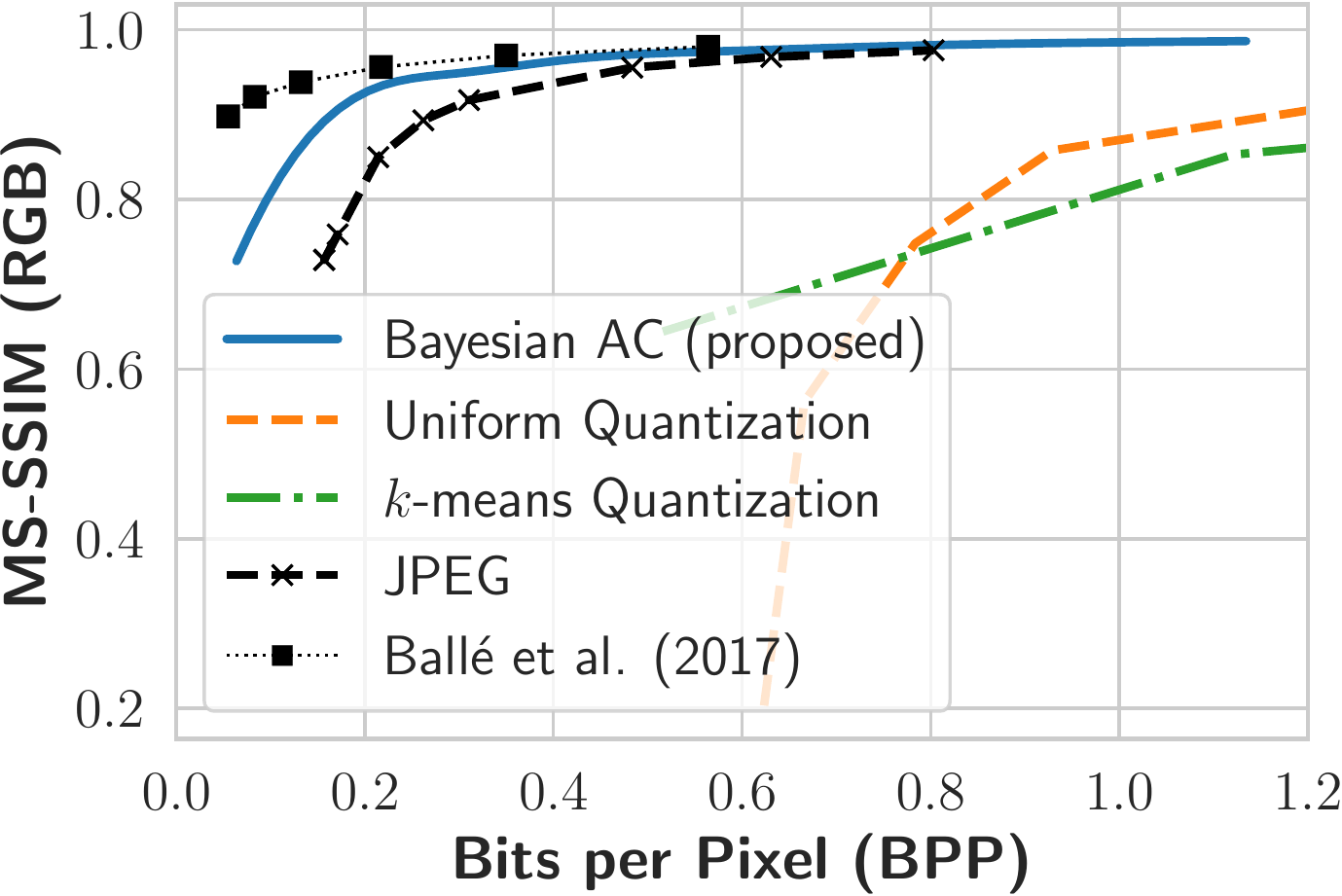}
 \end{subfigure}

 \vspace{10pt}\includegraphics[width=0.7\textwidth]{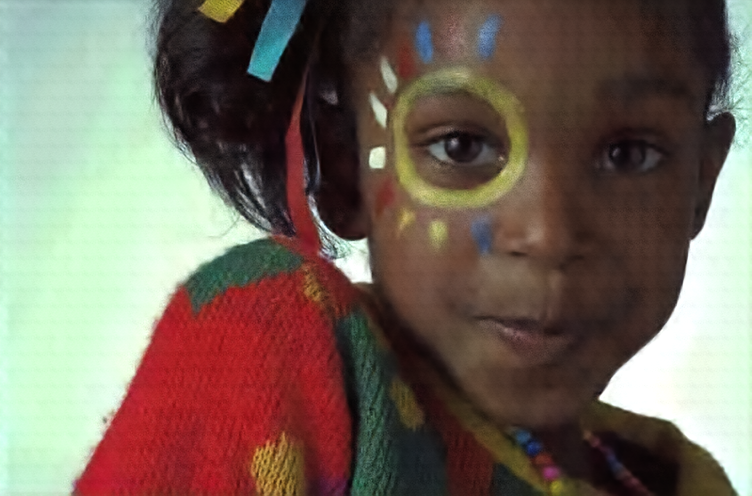}\vspace{-10pt}\caption{Proposed. bits-per-pixel: 0.2, PSNR: 29.523, MS-SSIM: 0.928}
 \vspace{10pt}\includegraphics[width=0.7\textwidth]{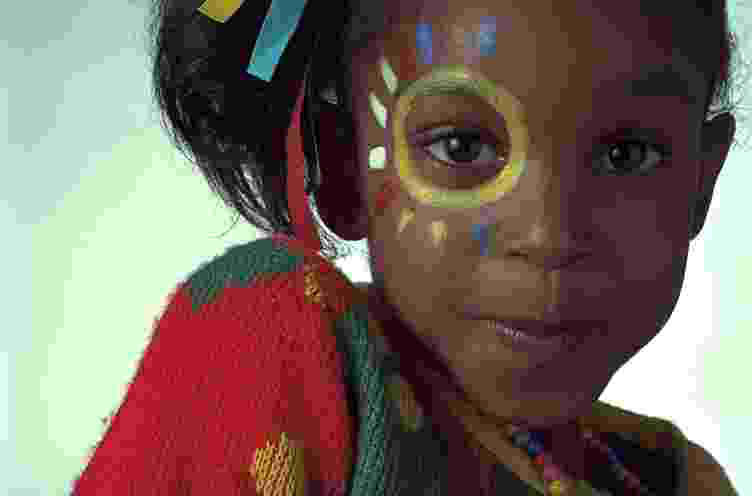}\vspace{-10pt}\caption{JPEG. bits-per-pixel: 0.2, PSNR: 26.6, MS-SSIM: 0.838}
\end{figure}

\begin{figure}[t]
\centering
 \begin{subfigure}[h]{0.45\columnwidth}
 \centering
 \includegraphics[width=\textwidth]{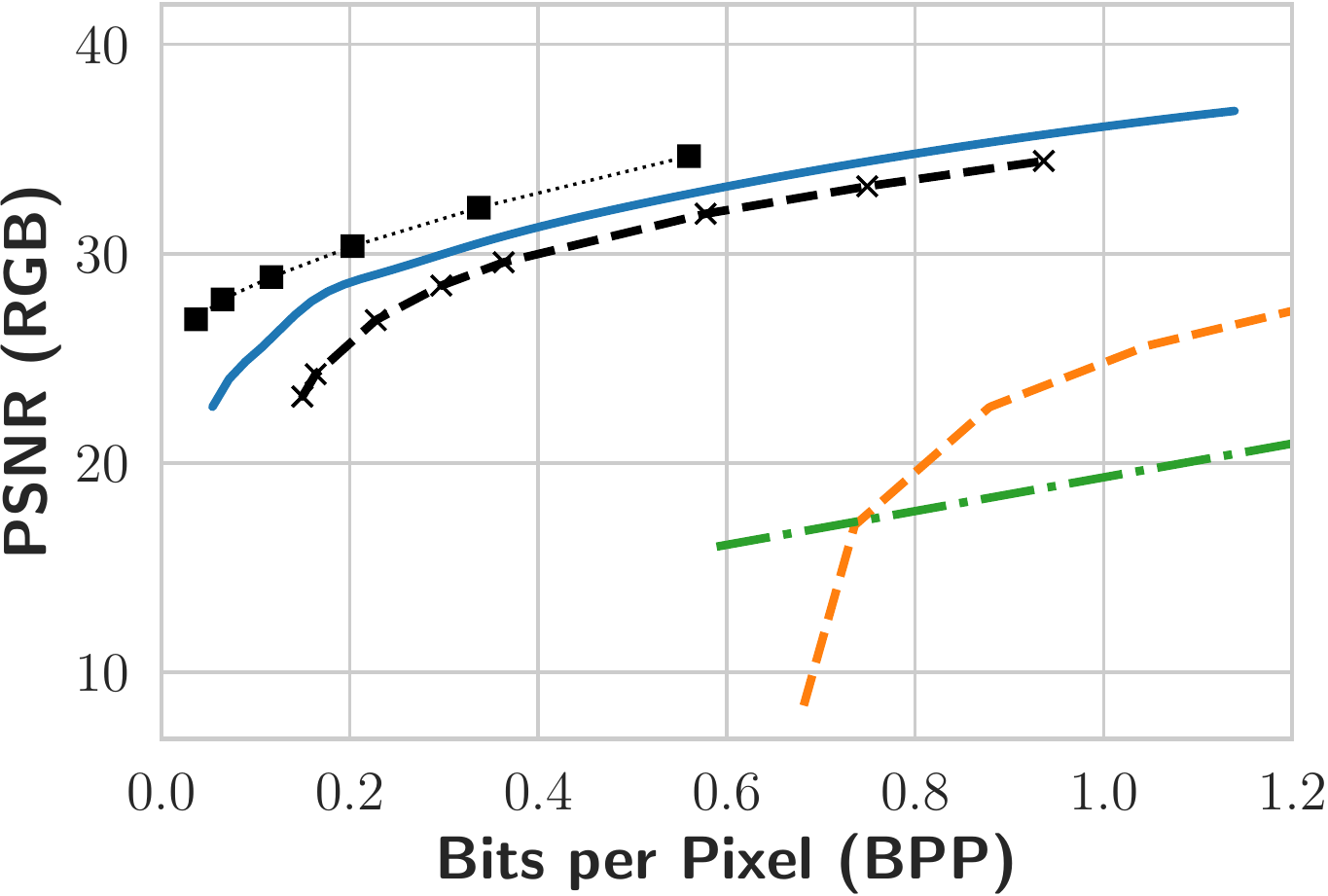}
 \end{subfigure}
 \hfill
 \begin{subfigure}[h]{0.45\columnwidth}
 \centering
 \includegraphics[width=\textwidth]{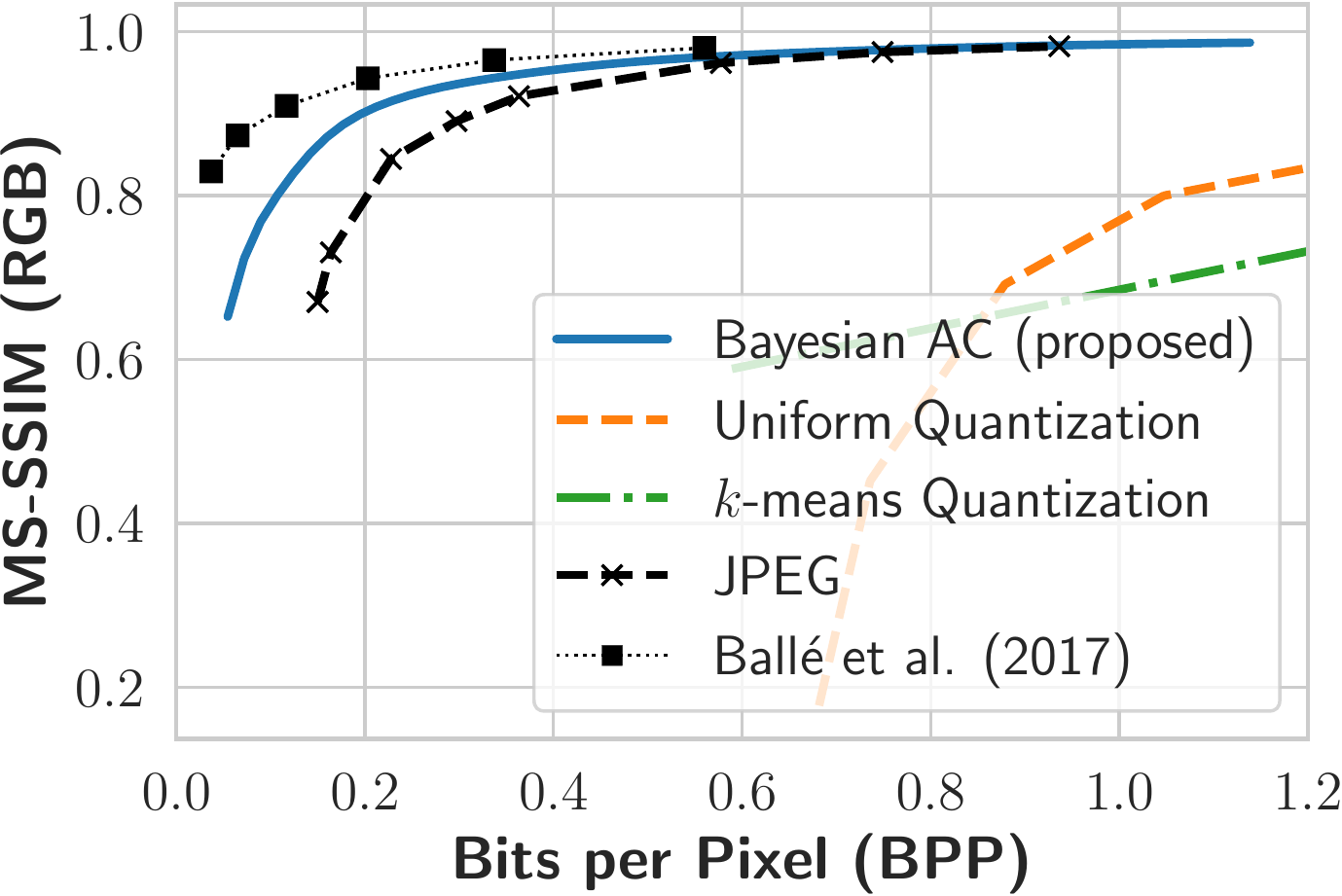}
 \end{subfigure}

 \vspace{10pt}\includegraphics[width=0.7\textwidth]{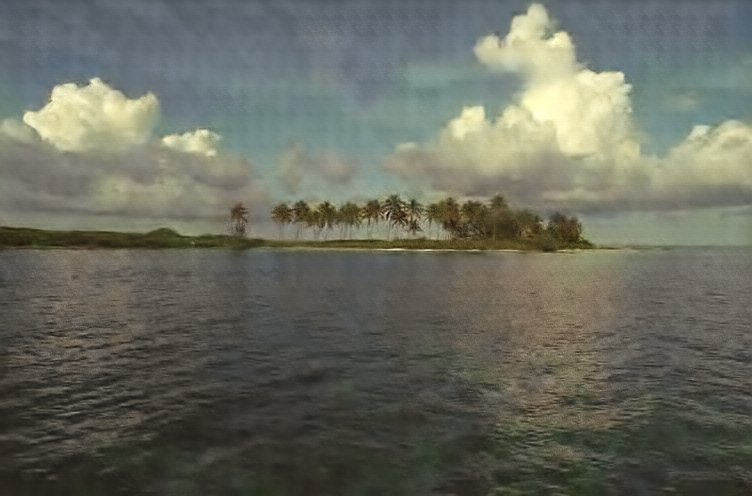}\vspace{-10pt}\caption{Proposed. bits-per-pixel: 0.21, PSNR: 28.77, MS-SSIM: 0.908}
 \vspace{10pt}\includegraphics[width=0.7\textwidth]{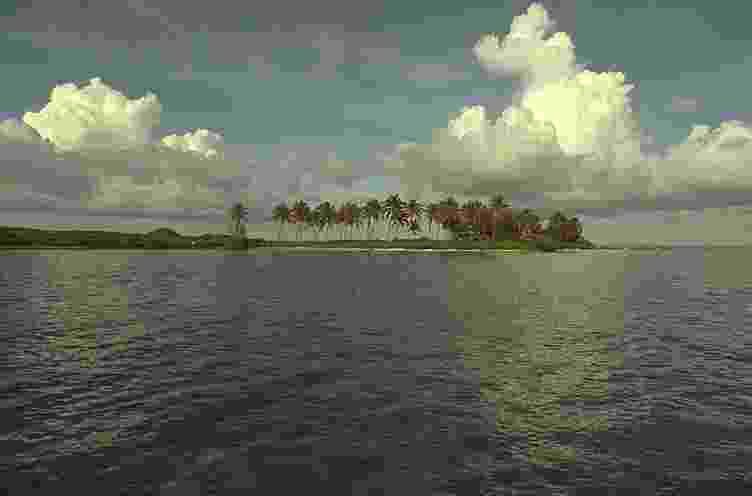}\vspace{-10pt}\caption{JPEG. bits-per-pixel: 0.21, PSNR: 26.247, MS-SSIM: 0.818}
\end{figure}

\begin{figure}[t]
\centering
 \begin{subfigure}[h]{0.45\columnwidth}
 \centering
 \includegraphics[width=\textwidth]{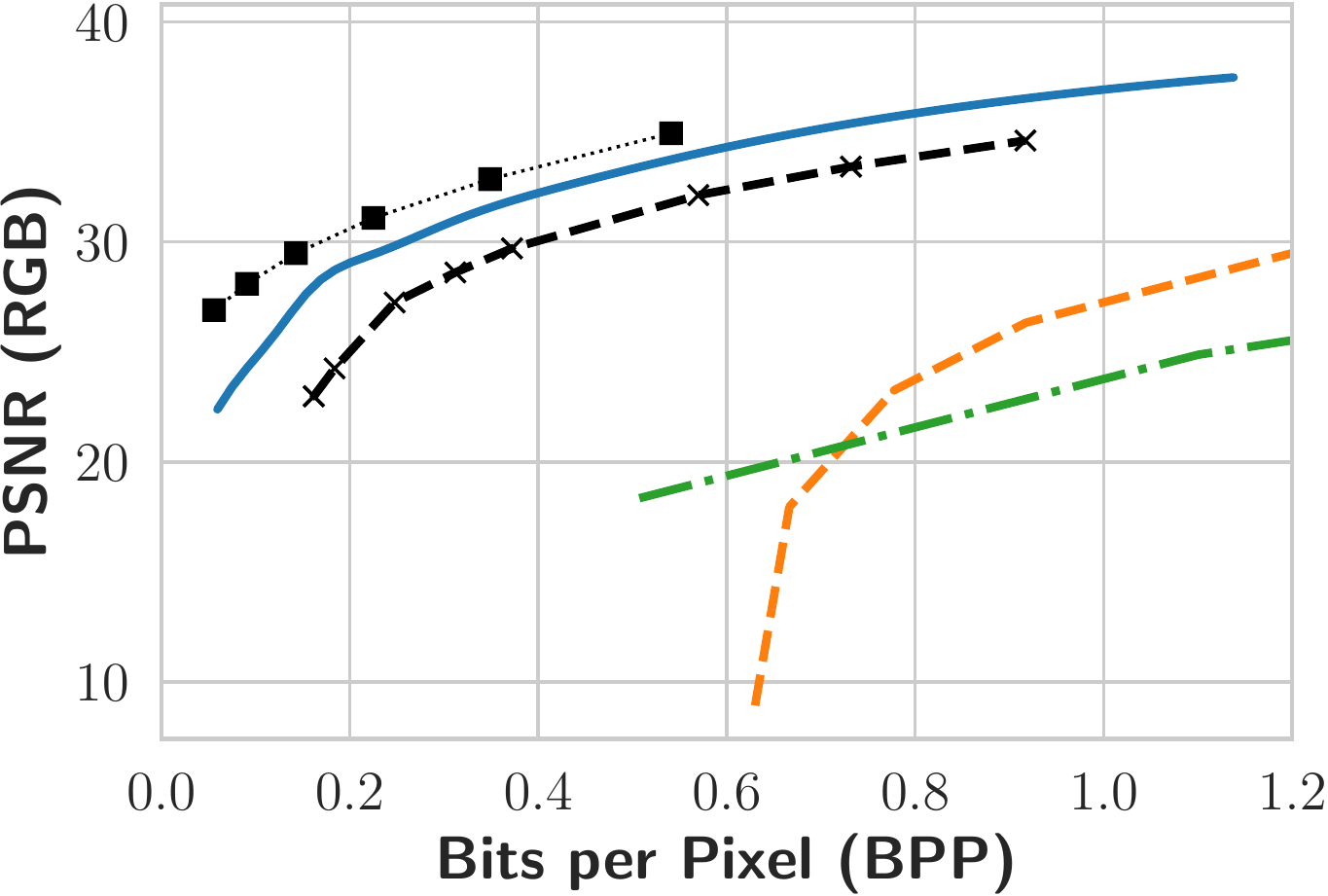}
 \end{subfigure}
 \hfill
 \begin{subfigure}[h]{0.45\columnwidth}
 \centering
 \includegraphics[width=\textwidth]{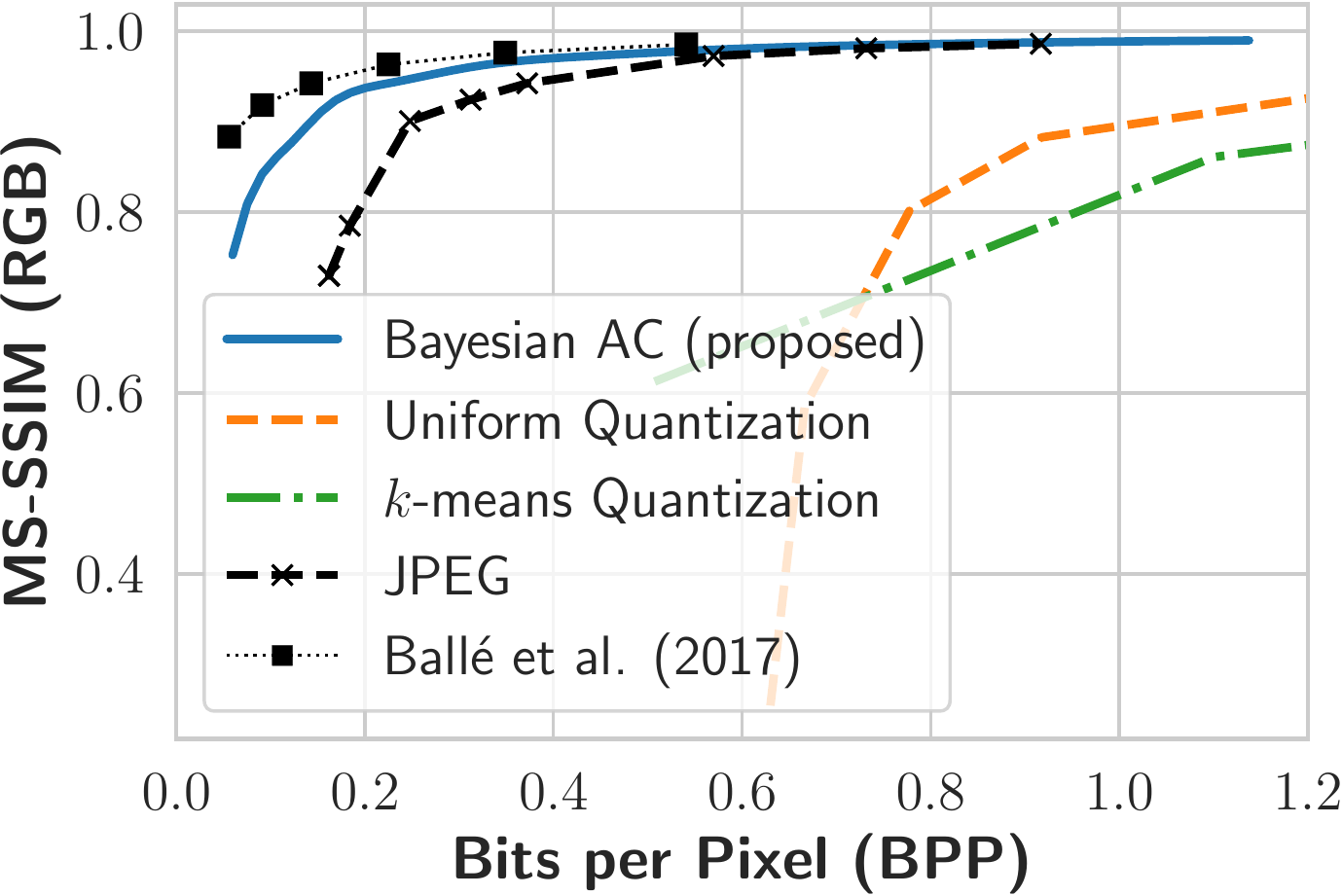}
 \end{subfigure}

 \vspace{10pt}
 \begin{subfigure}[h]{0.49\columnwidth}
 \centering
 \includegraphics[width=\textwidth]{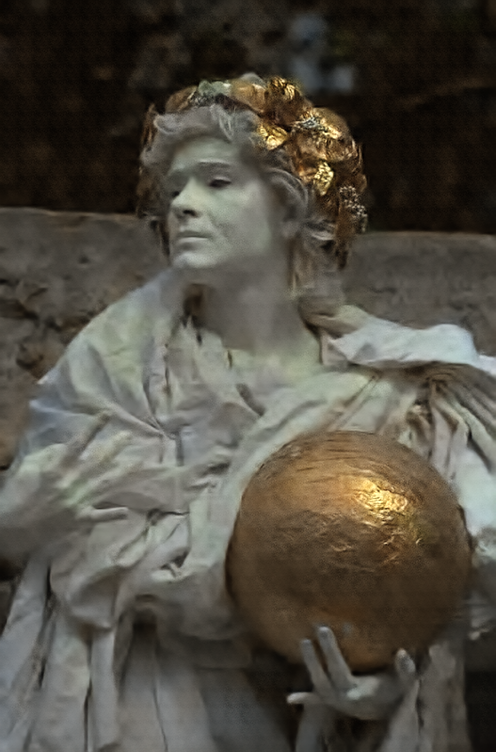}\caption{Proposed. bits-per-pixel: 0.2, PSNR: 29.006, MS-SSIM: 0.936}
 \end{subfigure}
 \hfill
 \begin{subfigure}[h]{0.49\columnwidth}
 \centering
 \includegraphics[width=\textwidth]{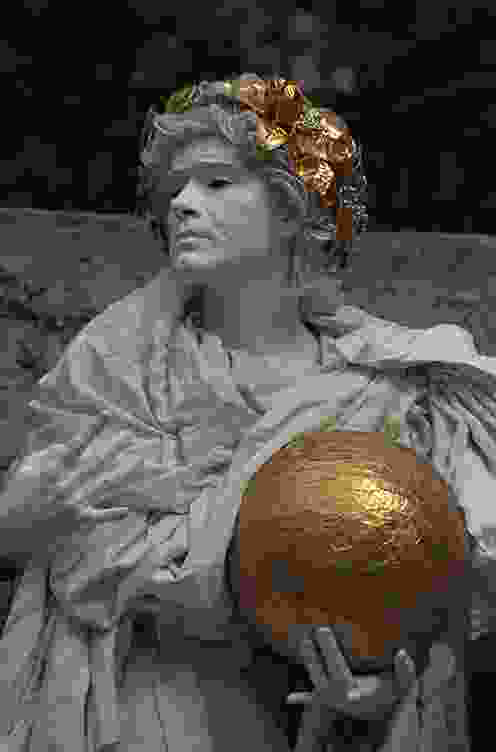}\caption{JPEG. bits-per-pixel: 0.2, PSNR: 25.389, MS-SSIM: 0.835}
 \end{subfigure}
\end{figure}

\begin{figure}[t]
\centering
 \begin{subfigure}[h]{0.45\columnwidth}
 \centering
 \includegraphics[width=\textwidth]{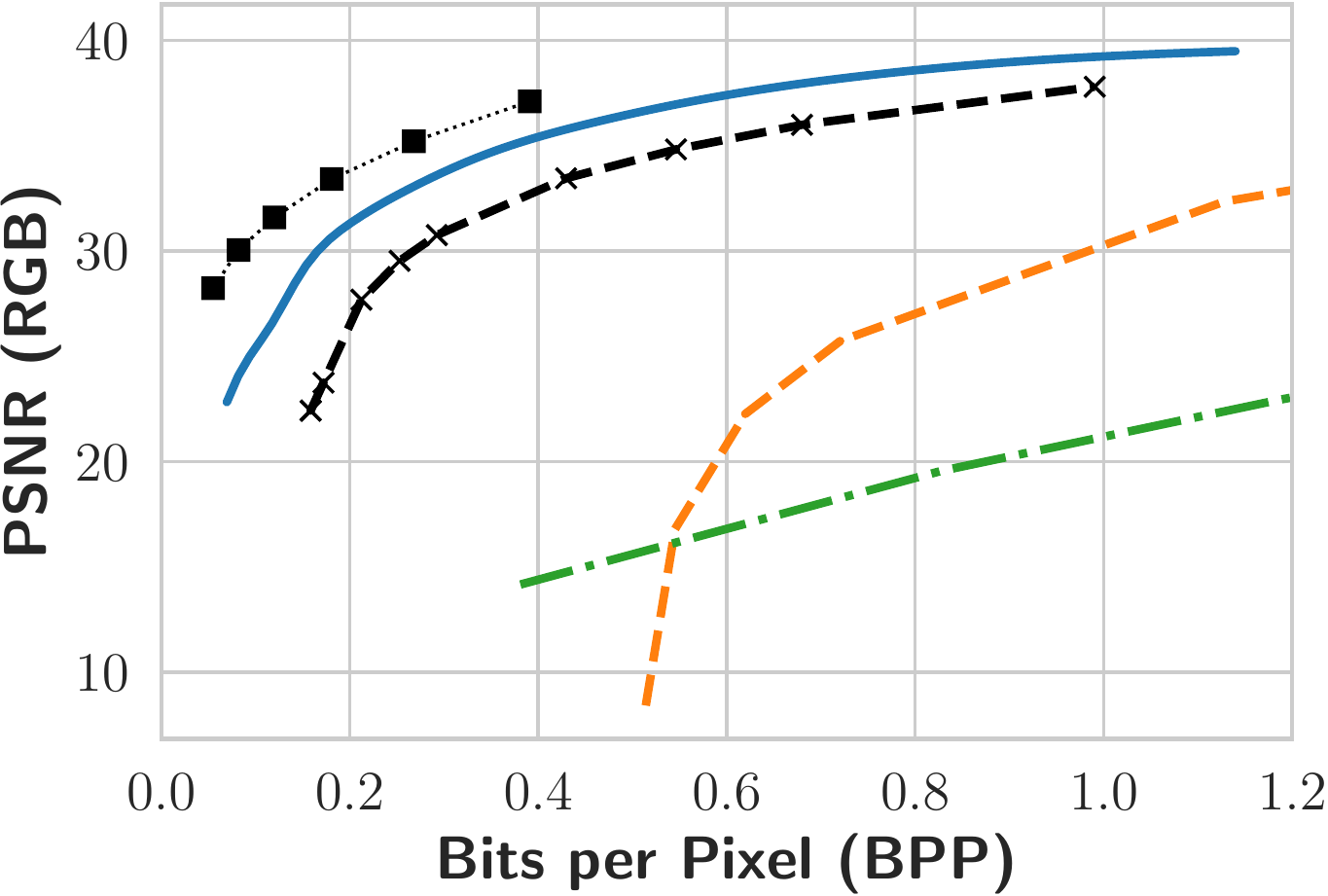}
 \end{subfigure}
 \hfill
 \begin{subfigure}[h]{0.45\columnwidth}
 \centering
 \includegraphics[width=\textwidth]{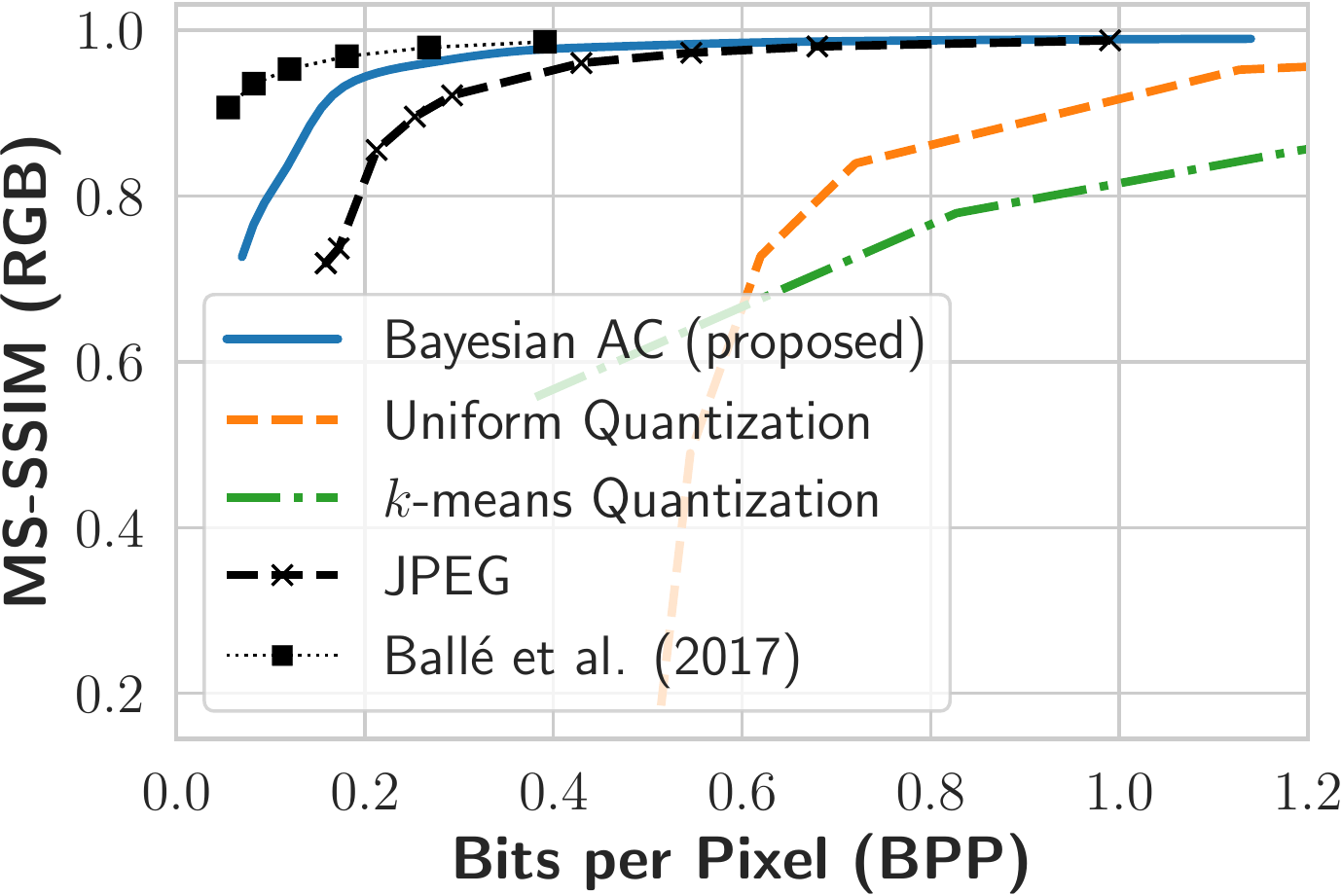}
 \end{subfigure}

 \vspace{10pt}\includegraphics[width=0.7\textwidth]{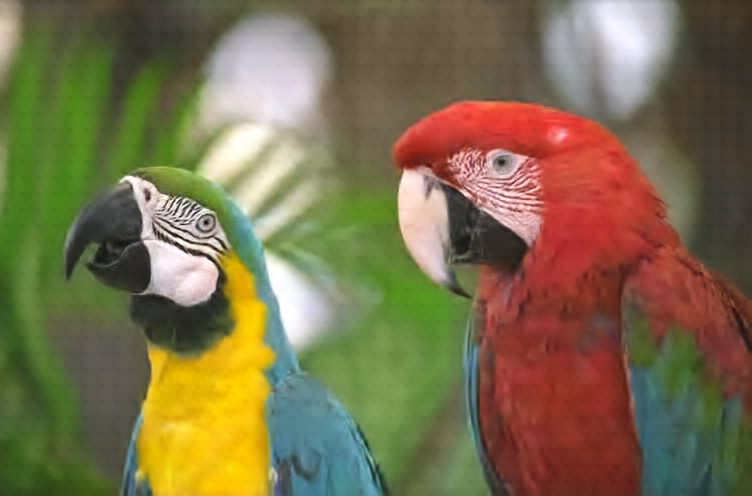}\vspace{-10pt}\caption{Proposed. bits-per-pixel: 0.2, PSNR: 31.425, MS-SSIM: 0.945}
 \vspace{10pt}\includegraphics[width=0.7\textwidth]{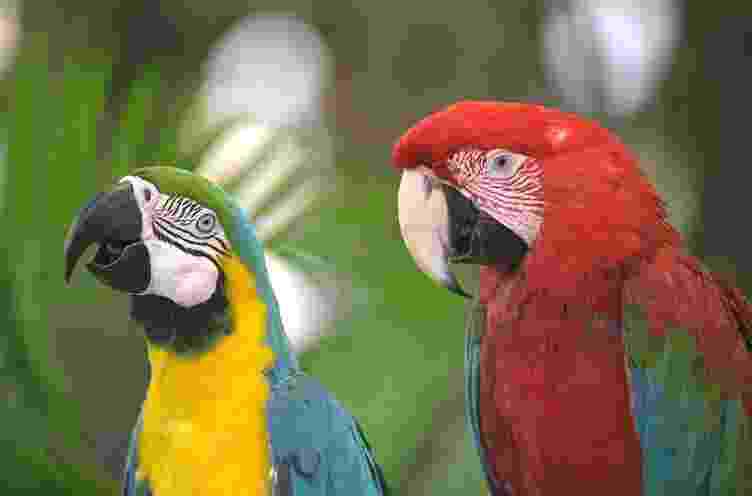}\vspace{-10pt}\caption{JPEG. bits-per-pixel: 0.2, PSNR: 26.976, MS-SSIM: 0.833}
\end{figure}

\begin{figure}[t]
\centering
 \begin{subfigure}[h]{0.45\columnwidth}
 \centering
 \includegraphics[width=\textwidth]{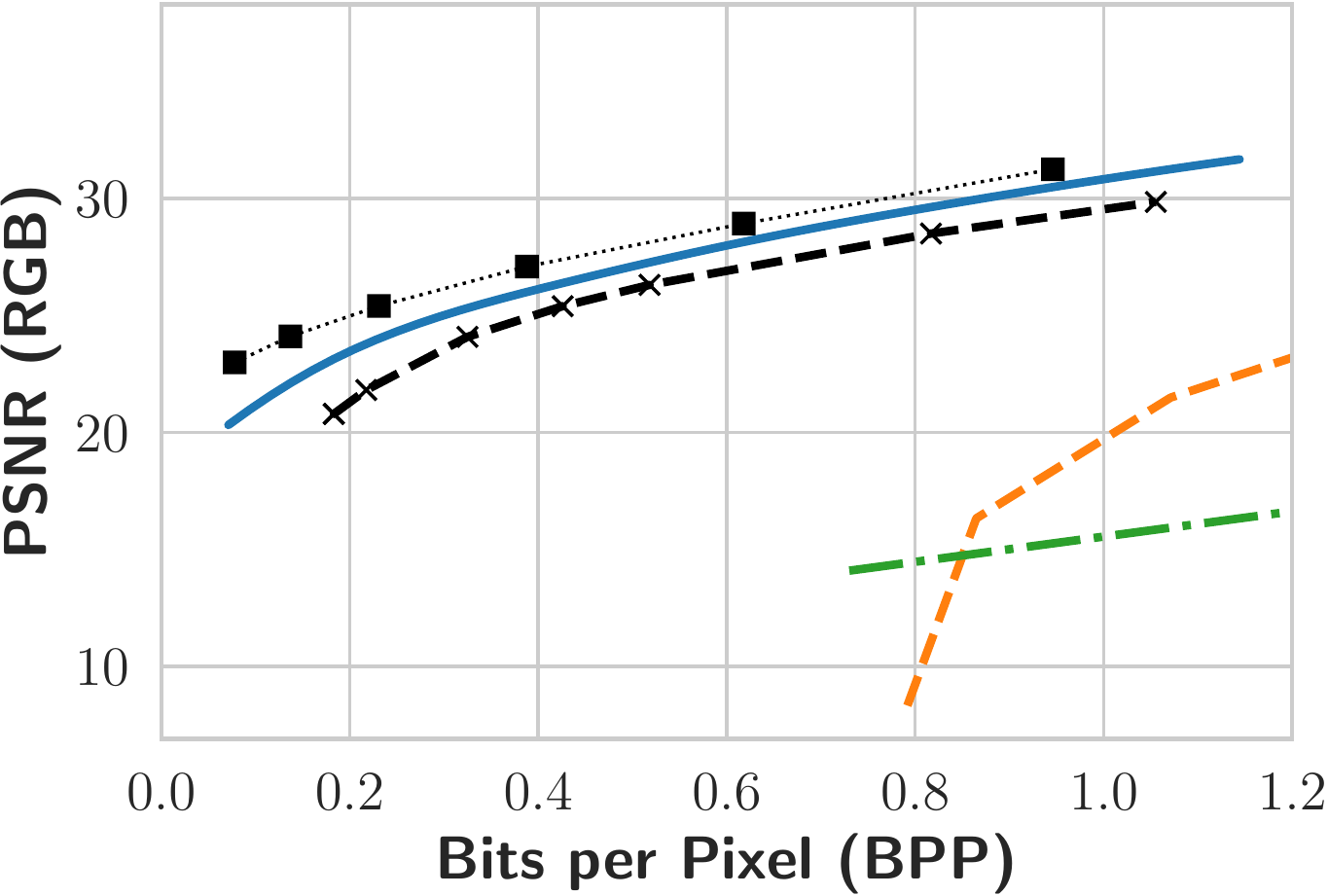}
 \end{subfigure}
 \hfill
 \begin{subfigure}[h]{0.45\columnwidth}
 \centering
 \includegraphics[width=\textwidth]{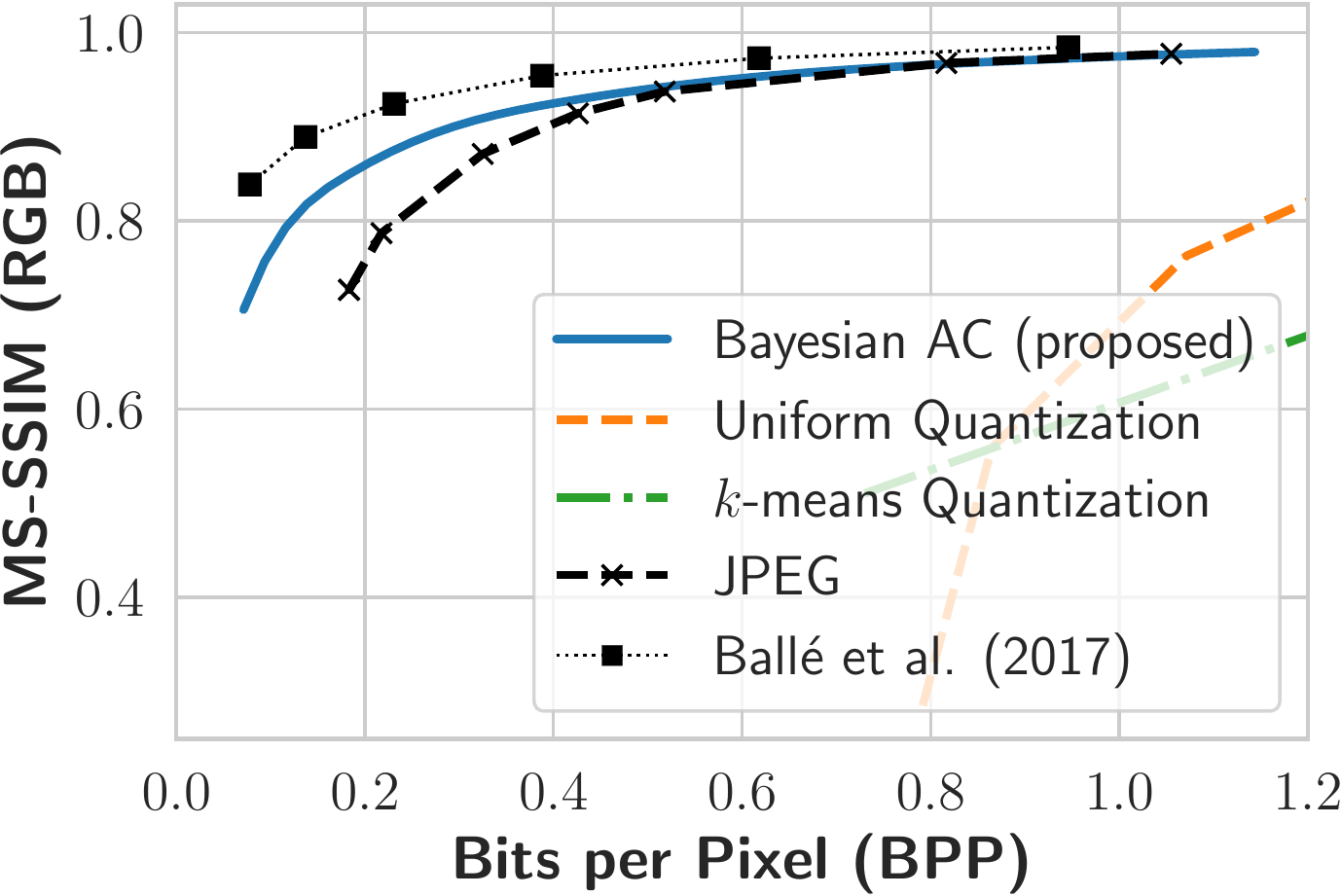}
 \end{subfigure}

 \vspace{10pt}\includegraphics[width=0.7\textwidth]{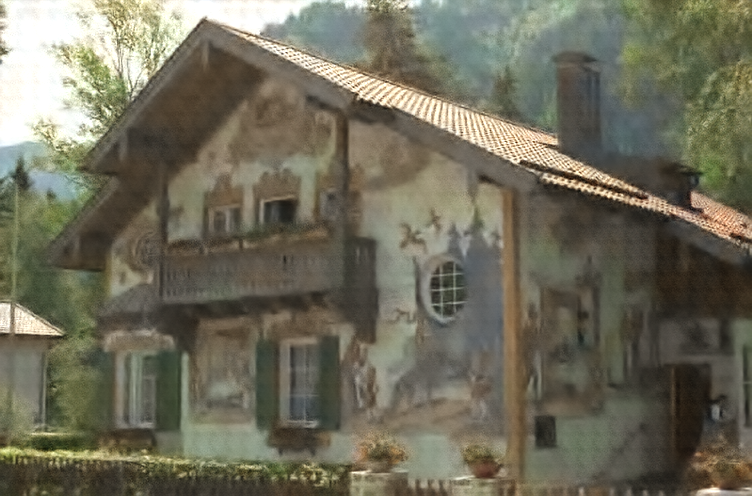}\vspace{-10pt}\caption{Proposed. bits-per-pixel: 0.24, PSNR: 24.254, MS-SSIM: 0.882}
 \vspace{10pt}\includegraphics[width=0.7\textwidth]{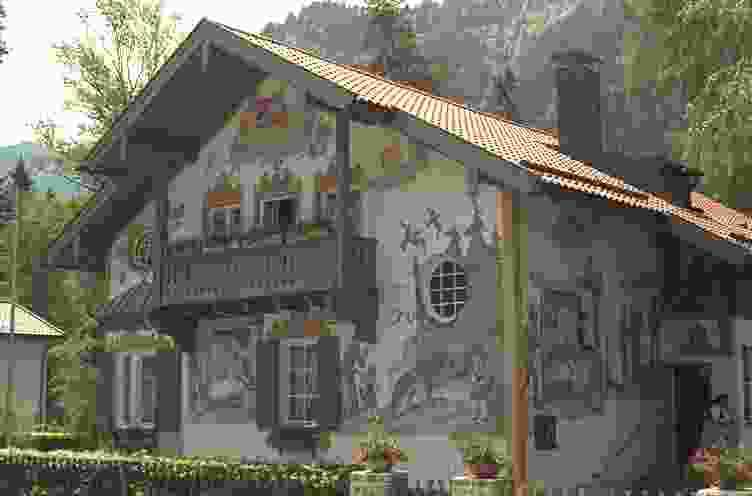}\vspace{-10pt}\caption{JPEG. bits-per-pixel: 0.24, PSNR: 22.562, MS-SSIM: 0.818}
\end{figure}